\documentclass[12pt]{article}
\usepackage{amsmath}
\usepackage{enumerate}
\usepackage{enumitem}
\usepackage[ruled,vlined]{algorithm2e}
\usepackage[hang,small,tight]{subfigure} 
\usepackage[utf8]{inputenc} 
\usepackage[T1]{fontenc}    
\usepackage{lmodern}
\usepackage{hyperref}       
\usepackage{url}            
\usepackage{booktabs}       
\usepackage{amsfonts}       
\usepackage{nicefrac}       
\usepackage{microtype}      
\usepackage{amsbsy}   
\usepackage{natbib}   
\usepackage{cleveref}       
\usepackage{doi}
\usepackage{graphicx}   
\usepackage{longtable}
\usepackage{pdfpages}
\usepackage{multirow}
\usepackage{color}   
\usepackage{indentfirst}   
\usepackage{tikz}
\usepackage{apalike}
\usetikzlibrary{matrix}
\usepackage{here}
\usepackage{appendix}
\usepackage{fixmath}
\usepackage{float}
\usepackage{titlesec} 
\usepackage{ulem}
\usepackage{bm} 
\usepackage{lscape} 
\usepackage{caption}
\usepackage{threeparttable} 
\usepackage{comment}
\usepackage{xr}
\usepackage{abstract}

\makeatletter

\newcommand{\bd}{\textbf}

\usepackage[all]{hypcap}  

\usepackage{titletoc}
\newcommand{\PR}[1]{\ensuremath{\left[#1\right]}}
\newcommand{\PC}[1]{\ensuremath{\left(#1\right)}}
\newcommand{\chav}[1]{\ensuremath{\left\{#1\right\}}}

\newcommand{\blind}{0}

\newcommand{\nofig}{0}

\usepackage{setspace}
\begin{document}

\doublespacing




\if0\blind
{\title{ An Efficient Sequential Approach for $k$-Parametric Dynamic Generalized Linear Models}
\vspace{-10pt}
  \author{Mariane B. Alves\footnotemark[1]  \footnotemark[2] , 
  Helio S. Migon\footnotemark[1], 
  Silvaneo V. dos Santos Jr.\footnotemark[1], 
  Raíra Marotta\footnotemark[1] \vspace{-10pt}\\
 \footnotemark[1] \footnotesize{Instituto de Matemática, Universidade Federal do Rio de Janeiro, Rio de Janeiro, Brazil.}\vspace{-10pt}\\
 \footnotemark[2] \footnotesize{Address for correspondence: Mariane B. Alves, Universidade Federal do Rio de Janeiro.}\vspace{-15pt}\\ \footnotesize{Av. Athos da Silveira Ramos, 149 - Centro de Tecnologia, Bl C (Térreo),}\vspace{-15pt}\\ \footnotesize{Rio de Janeiro, 21941-909, Brazil. E-mail: mariane@im.ufrj.br.}
    }

  \maketitle
} \fi
\if1\blind
{
  \bigskip
  \bigskip
  \bigskip
  \begin{center}
    {\bf An Efficient Sequential Approach for $k$-parametric Dynamic Generalized Linear Models}
\end{center}
} \fi
\date{}
\begin{abstract}
This paper introduces a novel sequential inferential method for Bayesian dynamic generalized linear models (DGLM), extending them to  $k$-parametric exponential families and accommodating both univariate and multivariate responses. The method efficiently handles various response types, including multinomial, gamma, normal, and Poisson distributions, leveraging the conjugate and predictive structures of the exponential family. It enables the assignment of dynamic predictors to the $k$ parameters that govern the response. By incorporating information geometry concepts, such as the projection theorem and Kullback-Leibler divergence, it aligns with recent variational inference advances, with the advantage of properly addressing joint uncertainty. Applications to synthetic and real datasets highlight its computational efficiency and scalability, surpassing methods like Markov Chain Monte Carlo, Variational Bayes, Integrated Nested Laplace Approximation and Particle Filter. The R package \texttt{kDGLM}, developed in conjunction with this investigation, is available for applied researchers, enabling easy implementation of the proposed method for
$k$-parametric dynamic generalized models.
\end{abstract}

\noindent%
\vfill

\newpage
\section{Introduction}

Modeling time series with multiparametric Gaussian and non-Gaussian distributions remains a central challenge in applied statistics, epidemiology, econometrics, and related fields. Over the last decades, substantial advances have been achieved through \textit{dynamic generalized linear models} (DGLMs) \citep{west1985dynamic}, which extend both generalized linear models (GLMs, \citealp{nelder1972generalized}) and dynamic linear models \citep{Harrison1976bayesian}. DGLMs provide a Bayesian framework for modeling time-dependent responses from the \textit{uniparametric exponential family}, enabling sequential updating and online forecasting. However, their  Bayesian formulation is restricted to \textit{single-parameter} exponential families. This limitation precludes direct inference for more complex $k$-parametric ($k>1$) distributions—such as bivariate normal, gamma, or multinomial models—whose parameters evolve jointly over time.

In many modern applications, such as  dynamic compositional data or macroeconomic analysis, stochastic volatility modeling,  single-parameter models are insufficient. For instance, in econometrics, researchers are often interested in the joint evolution of the \textit{mean and volatility} of key indicators. Recent examples include \citet{Alves2025}, that specified a bivariate dynamic normal model for output gap and inflation, assigning dynamics to the mean and log-precision of each indicator, as well as to their mutual correlation, incorporating dynamic trends, seasonality,  and transfer effects. These developments highlight the growing need for flexible frameworks that accommodate multiple evolving parameters within a coherent sequential Bayesian structure.


This paper addresses the problem of extending dynamic models to the \textit{$k$-parametric exponential family}, where the observed process $y_t$, $t = 1, 2, \ldots$, is conditionally independent given latent states $\theta_t$ that evolve stochastically over time. Although Bayesian updating for Gaussian models is analytically tractable \citep{Migon23, west1997}, non-Gaussian dynamic models often require computationally intensive simulation-based methods, such as MCMC \citep{ fruhwirth1994data,Gamerman1998Markov}. Sequential Monte Carlo (SMC) methods \citep{Gordon1993Bootstrap, Doucet2001SMCBook} and their variants---including sequential MCMC \citep{Andrieu2010PMCMC}, ensemble Kalman and Gaussian mixture filters \citep{Evensen1994EnKF, Reich2012EGMF}, and the Liu--West filter \citep{Liu2016KernelEnGMF}---offer alternatives but often suffer from poor scalability in high dimensions and loss of analytical tractability.


We propose a general extension of DGLMs---denoted as \textit{kDGLM}---that allows sequential Bayesian inference for univariate or multivariate $k$-parametric exponential family models. The framework supports dynamic trends, seasonality, and regression effects, while maintaining analytical sequential updating and computational efficiency. It thus integrates the advantages of traditional DGLMs. The inferential procedure is derived from \textit{information geometry} \citep{amari2016information}, specifically using \textit{Bregman’s projection theorem} to minimize the divergence between the canonical parameter conjugate prior and the state-induced prior. This leads to a variationally inspired approximation---similar in spirit to Expectation Propagation \citep{Vehtari2020}  and Variational Bayes \citep{Kucukelbir2017,Blei2017, Koop2023}---but with the crucial distinction that our method provides adequate quantification of joint uncertainty, is fully sequential and analytically tractable. Moreover, the proposed structure resolves the imbalance between the $k$-dimensional parameter vector and the $p$-dimensional latent state vector, a longstanding limitation of classical DGLMs \citep{west1985dynamic, souza2016extended}.

Our approach is computationally efficient by design. Since updates are sequential, total cost grows linearly with the number of observations and the memory footprint is effectively constant, as each datum can be discarded once assimilated. As time and memory growth are central bottlenecks in modern Bayesian inference \citep{Guhaniyogi03072018}, this design targets both directly. Although the updates use analytic approximations, in our experiments the resulting inferences attain accuracy comparable to both sequential and non-sequential MCMC baselines  - particle filter \citep{Liu2016KernelEnGMF} and No-U-Turn Sampler \citep[NUTS, ][]{Hoffman2014,STAN2017}, while requiring substantially less computation than approximate alternatives  like Integrated Nested Laplace Approximation \citep[INLA, ][]{Rue2009INLA,Rue2017} and Variational Bayes \citep{Blei2017}. As shown across the considered datasets and illustrations, we observe up to 7× lower wall-clock time than the most competitive approximate method and more than two orders of magnitude savings relative to non-sequential baselines.


The main contributions of this work can be summarized as follows:

\begin{enumerate}
    \item a unified \textit{kDGLM framework} is developed for Bayesian sequential analysis of dynamic models within the $k$-parametric exponential family, supporting multiple evolving parameters;
    \item an analytically tractable approximation method derived from \textit{information geometry} ensures efficient online inference without simulation;
    \item computational scalability and preservation of sequential updating, enabling online monitoring, forecasting, and intervention---key advantages over particle filtering or MCMC-based approaches;
    \item practical implementation through the \texttt{kDGLM} R package \citep{santos2024}, facilitating adoption by applied researchers.
\end{enumerate}



The remainder of the paper is organized as follows. Section~\ref{sec:basic} reviews the background of DGLMs and alternative inferential approaches. Section~\ref{sec:OurProp} presents the principles of information geometry that are applied and details of the proposed \textit{kDGLM} algorithm. Sections~\ref{case} and ~\ref{illust} provide a range of applications using synthetic and real datasets, including multinomial, normal, gamma, and Poisson dynamic models.  Finally, Section~\ref{concl} concludes with remarks on possible extensions and future research directions.

\section{A Review of Dynamic Generalized Linear Models}
\label{sec:basic}

In this section, we review key concepts, underlying the models used throughout this paper. We focus on uni- or multivariate  time-indexed observables $ \mathbf{y}_t $, $ t=1, 2, \ldots $, which follow a $k$-parametric exponential family distribution ($k \ge 1$) and are assumed to be conditionally independent given the $ k $-dimensional parameter vector $ \bm{\eta}_t $. The probability density (or mass) function of the observables is expressed as $ p(\mathbf{y}_t| \bm{\eta}_t) = f(\mathbf{y}_t)a( \bm{\eta}_t ) \exp \left\{ \sum_{i=1}^k c_i \phi_i(\bm{\eta}_t) h_i(\mathbf{y}_t) \right\} $, which can be rewritten in canonical form as:
\begin{equation}
\label{eqfamexp}
    p(\mathbf{y}_t|\bm{\psi}_t) = c(\mathbf{y}_t)\exp\chav{{\mathbf{H}}'(\mathbf{y}_t)\bm{\psi}_t - b(\bm{\psi}_t)}, 
\end{equation}
where $ {\mathbf{H}}(\mathbf{y}_t)= (h_1(\mathbf{y}_t), \ldots, h_k(\mathbf{y}_t))'$ is a sufficient statistic vector matching the dimension of the parameter vector $\bm{\eta}_t$, and $\bm{\psi}_t$ is the canonical parameter at time $t$, given by $\bm{\psi}_t = (\psi_{1t}, \ldots, \psi_{kt})'$ and $\psi_{it} = c_i\phi_i(\bm{\eta}_t)$ for $i = 1, \ldots, k$. 

The conjugate prior distribution $CP(\tau_0,\bm{\tau})$ for $\bm{\psi}_t$ is expressed as
$
p(\bm{\psi}_t|\bm{\tau}) = [K(\tau_0,\bm{\tau})]^{-1}\,\,\\ \exp[\bm{\tau}'\bm{\psi}_t - \tau_0 b(\bm{\psi}_t)],
$
where $\bm{\tau}' = (\tau_1, \ldots, \tau_k)$. It is well established that if $\mathbf{y}_t|\psi_t$ follows a distribution in the exponential family, then $E[{\mathbf{H}}(\mathbf{y})] = \nabla b(\bm{\psi})$ and $V[{\mathbf{H}}(\mathbf{y})] = \nabla^2 b(\bm{\psi})$, where $\nabla$ is the differential operator. For regular exponential families, under conjugate priors, posterior and predictive distributions are analytically tractable \citep{bernardo2001bayesian}, except when the normalizing constant of the conjugate prior is unknown. Specifically, the posterior density of $\bm{\psi}_t$ given a new observation $\mathbf{y}_t$ at time $t$ is $p(\bm{\psi}_t|\mathbf{y}_t,\tau_0,\bm{\tau}) = p(\bm{\psi}_t|\tau_0^*, \bm{\tau}^*)$, where $\tau_0^* = \tau_0 + 1$ and $\bm{\tau}^* = \bm{\tau} + {\mathbf{H}}(\mathbf{y}_t)$, with $\bm{\tau} + {\mathbf{H}}(\mathbf{y}_t) = (\tau_1 + h_1(\mathbf{y}_t), \ldots, \tau_k + h_k(\mathbf{y}_t))'$. At each time $t$, the predictive distribution for a future observation $\mathbf{y}_f$ is given by $p(\mathbf{y}_f|\mathbf{y}_{1:t}, \tau_0, \bm{\tau}) = c(\mathbf{y}_f) 
{K(\tau_0^* + 1, \bm{\tau}^* + {\mathbf{H}}(\mathbf{y}_f))}/{K(\tau_0^*, \bm{\tau}^*)}$, where ${\mathbf{H}}(\mathbf{y}_f) = (h_1(\mathbf{y}_f), \ldots, h_k(\mathbf{y}_f))'$.

 \cite{west1985dynamic} introduces the DGLM class, focusing on sequential inference for uniparametric exponential families,  establishing a connection between parameter ${\eta}_t$ and a dynamic linear predictor $\mathbf{F}_t' \bm{\theta}_t$ via an invertible link function ${g}(\cdot)$. A conjugate prior CP($\tau_{0t},  \bm{\tau}_{t}$) is assigned to the canonical parameter ${\psi}_t = {c}{\phi}(\bm{\eta}_t)$ and the states $\bm{\theta}_t$ prior is partially specified, in terms of first and second moments. The $k$-parametric exponential family extension proposed in the present work considers a model defined by three components: an observable $k$-parametric exponential family model (see equation (\ref{eqfamexp})); a relation linking the $k$-dimensional vector $\bm{\eta}_t$ to a dynamic predictor vector $\bm{\lambda}_t$ governed by states $\bm{\theta}_t$ ($\bm{g}(\bm{\eta}_t) = \bm{F}_t'\bm{\theta}_t = \bm{\lambda}_t$); and an evolution equation for the latent states $\bm{\theta}_t = {\mathbf{G}}_t'\bm{\theta}_{t-1} + \bm{\omega}_t$ with $\bm{\omega}_t \sim N_p[\mathbf{0},\mathbf{W}_t]$. Here, ${\mathbf{F}}_t$ is a known ($p \times k$) dynamic regression matrix, ${\mathbf{G}}_t$ is a ($p \times p$) state evolution matrix, ${\mathbf{W}}_t$ is a ($p \times p$) evolution covariance matrix, and $\bm{g}$ is an invertible link function.
 This class provides conjugacy properties beneficial for sequential inference.

\cite{west1985dynamic}  resolved the dimensional imbalance between the $p$-dimensional state vector and the canonical parameter by employing linear Bayes estimation.   \cite{souza2016extended} introduced extended dynamic generalized linear models (EDGLMs) for the biparametric exponential family, building upon \cite{west1985dynamic}. EDGLMs employ a bivariate link function to relate linear predictors defined by states $\mathbold{\theta}_t$ to the canonical parameter vector $\mathbold{\eta}_t$. The sequential inference algorithm, based on conjugacy and linear Bayes estimation, encounters a challenge since the conjugate updating step can involve more conditions than unknowns. A solution inspired by the generalized method of moments was put forth to deal with this potential drawback.

In our study, we directly manage arbitrary dimensions of the parametric vector, thus avoiding the additional step introduced by \cite{souza2016extended}. By leveraging normal theory and properties of conditional expectation, we resolve issues arising from unbalanced dimensions between the canonical parameter and  state vectors. Additionally, by incorporating information geometry arguments, we extend the works of \cite{west1985dynamic} and \cite{souza2016extended}, offering a comprehensive formulation for sequential inference in $k$-parametric dynamic generalized linear models for $k \ge 1$.

\section{Sequential Inference in $k$DGLMs}\label{sec:OurProp}
We present a novel method for sequentially updating DGLMs for responses in the $k$-parametric exponential family, assuming conditionally independent observations $\mathbf{y}_t$ for $t=1,2,\ldots$, described by the observational model defined by equation (\ref{eqfamexp}). The $k$-dimensional vector of dynamic predictors, in turn, is defined by equation (\ref{eq.pred}) and guided  by latent states that evolve according to the  system (or evolution) equation (\ref{eq.sist}):
\begin{eqnarray} 
\bm{g}(\bm{\eta}_t) &=& \bm{F}_t'\bm{\theta}_t = \bm{\lambda}_t, \label{eq.pred}\\\bm{\theta}_t &=&  {\mathbf{G}}_t'\bm{\theta}_{t-1} + \bm{\omega}_t,\quad \bm{\omega}_t \sim N_p(\mathbf{0},\mathbf{W}_t), \label{eq.sist}
\end{eqnarray}
 with $\bm{F}_t$, ${\mathbf{G}}_t$ and $\mathbf{W}_t$ as defined in Section \ref{sec:basic}. Let $D_0$ represent the initial information set before any observation. The model incorporates a $p$-variate normal prior specification: $(\bm{\theta}_1|D_0) \sim N_p({\bm a}_1, {\bm R}_1)$. Discount factors are applied to implicitly specify ${\bm W}_t$ \citep[see][Chap. 6]{west1997}.

In what follows we provide a concise overview of the key concepts underlying our proposed algorithm for inference in the $k$-parametric uni- or multivariate DGLM. The core result is the {\it projection theorem} \citep[see][]{amari2016information}, which is applied to reconcile prior and posterior distributions derived from the state vector specification and the conjugate prior for the canonical parameters.

Let $\cal M$ be a manifold of probability distributions, ${\cal{M}} = \{ p( \bm{\eta}|\bm{\tau}), \, \bm{\tau} \in  \mathbf{E}  \subset \mathbb{R}^k \}$, where $p(\bm{\eta}|\bm{\tau})$ represents a probability density function and $\bm{\tau} \in \mathbf{E}$ serves as a coordinate system. Specifically, we focus on the manifold of the exponential family probability distributions, as defined in equation (\ref{eqfamexp}). For simplicity, we suppress temporal indices and assume $p(\bm{\eta} | \bm{\tau}) = c(\bm{\eta}) \exp[ \mathbf{H}'(\bm{\eta})\bm{\psi} - b(\bm{\psi})]$. Here, $b(\bm{\psi})$ denotes a convex cumulant generating function, and we examine the divergence associated with $b(\bm{\psi})$ \citep{amari2016information}.

The Bregman divergence is defined as $D_{b}[\bm{\psi} ; \bm{\psi}_0] = b(\bm{\psi}) - [ b(\bm{\psi}_0) + \nabla b(\bm{\psi}_0) \,  (\bm{\psi} -\bm{\psi}_0)].$ For the special case where $b(\bm{\psi})$  is the free energy convex function   of the exponential family \citep[][p.13]{amari2016information}, it results in the Kullback-Leibler (KL) divergence between distributions $p$ and $q$: 
$$
D_{KL}[p(\bm{\eta}|\bm{\tau}'); q(\bm{\eta}|\bm{\tau})] = \int p(\bm{\eta}|\bm{\tau}')\, \log\left(  \frac{p(\bm{\eta}|\bm{\tau}')}{q(\bm{\eta}|\bm{\tau})} \right) \, d\bm{\eta}.
$$

Our method to approximate priors and posteriors is a specific instance of Theorem 1.4 in \cite{amari2016information}. 

\newtheorem{theorem}{Theorem}[section]
\begin{theorem}[{\it Projection theorem}]\label{proj_theo}
 
Let $ p(\bm{\eta}) $ denote a probability distribution over a set $ \mathcal{Y} $ and let $ S $ represent an exponential family over $ \mathcal{Y} $. The distribution $ q(\bm{\eta}) $ within $ S $ that minimizes the Kullback-Leibler divergence $ D_{KL}(p \parallel q) $ satisfies $ E_q(\mathbf{H}_q) = E_p(\mathbf{H}_q) $, where $ \mathbf{H}_q $ denotes the vector of sufficient statistics under $ q $ and $E_p$, $E_q$, respectively, denote expectations under $p$ and $q$. 
\end{theorem} The proof of Theorem \ref{proj_theo} is straightforward.\\

\medskip

\centerline{\bf Sequential learning in $k$DGLMs}

\medskip

The method of \textit{conjugate updating} proposed by \cite{west1985dynamic}, as briefly outlined in Section \ref{sec:basic}, characterizes the probabilistic nature of states $\bm{\theta}_t$ through their first and second moments, then equalizing the resulting moments of the linear predictors $\bm{\lambda}_t$ to the ones of the conjugate prior for  $\bm{g}(\bm{\eta}_t)$. In contrast to restricting reconciliation to moment-based specifications, we propose reconciling entire prior and posterior densities.  In our proposal, we assume Gaussian prior densities for the states $\bm{\theta}_t$ resulting in a $k$-dimensional Gaussian distribution for the linear predictor vector $\bm{\lambda}_t$. Then we minimize the Kullback-Leibler divergence between this Gaussian prior and the joint conjugate prior for $\bm{\lambda}_t = \bm{g}(\bm{\eta}_t)$.   Reconciling full density curves is more discerning than matching moments, since proximity in the parametric space (or between parameter estimates) may not reflect how similar two alternative prior (or posterior) densities are. Although this involves minimizing the divergence between two densities, the problem ultimately reduces to an optimization task within the parametric space. At each time $t$, we aim to find the parameters $\bm{\tau}_t$ of the conjugate prior for $\bm{\eta}_t$ that minimize the KL divergence from the prior induced by the Gaussian specification for the states.

Once the conjugate prior is fully specified and $\mathbf{y}_t$ is observed, updating is executed via Bayes' theorem. This yields a posterior distribution within the same parametric family as the prior, with updated parameters $\bm{\tau}^*_t$. Our objective, then, is to minimize the KL divergence between the posterior distribution of the linear predictors, as implied by the conjugate posterior, and a $k$-variate normal distribution for the same object, $\bm{\lambda}_t$. Given that $\bm{\lambda}_t$ is a linear transformation of $\boldsymbol{\theta}_t$, and assuming prior normality of the states, $(\boldsymbol{\theta}_t, \bm{\lambda}_t | D_{t-1})$ follows a Gaussian distribution, where $D_{t-1} = \{D_0, \mathbf{y}_1, \ldots, \mathbf{y}_{t-1}\}$. The updated distribution of $\boldsymbol{\theta}_t$ depends on $\mathbf{y}_t$ only through $\bm{\lambda}_t$. Consequently, normal theory provides a closed-form analytical solution for the conditional distribution $(\boldsymbol{\theta}_t | \bm{\lambda}_t, D_{t-1})$. The distribution of $(\bm{\lambda}_t | D_t)$ is thus approximated by a normal specification, and  thus the first and second moments of $(\boldsymbol{\theta}_t | D_t)$ are determined as in \citet[p.639]{west1997}. This posterior distribution at time $t$ evolves to a prior specification for $\bm{\theta}_{t+1}$, and the learning cycle repeats.

Algorithm \ref{algo:b_v2} outlines the information filtering process in our method. We use the notation established in \cite{west1997} for the mean vectors and covariance matrices:

i) posterior distribution at time $ t-1 $: $ \mathbf{m}_{t-1} = E[\bm{\theta}_{t-1} \mid D_{t-1}] $ and $ \mathbf{C}_{t-1} = \text{Cov}[\bm{\theta}_{t-1} \mid D_{t-1}] $;

ii) prior distribution for time $ t $: $ \boldsymbol{a}_{t} = E[\bm{\theta}_t \mid D_{t-1}] $ and $ \mathbf{R}_{t} = \text{Cov}[\bm{\theta}_t \mid D_{t-1}] $;

iii) one-step-ahead predictive distribution of the linear predictor: $ \bm{f}_t = E[\bm{\lambda}_t \mid D_{t-1}] $ and $ \bm{Q}_t = \text{Cov}[\bm{\lambda}_t \mid D_{t-1}] $;

iv) posterior distribution of the linear predictor at time $ t $: $ \bm{f}^*_t = E[\bm{\lambda}_t \mid D_{t}] $ and $ \bm{Q}^*_t = \text{Cov}[\bm{\lambda}_t \mid D_{t}] $.

Algorithm \ref{algo:b_v3} details the steps involved in smoothing and prediction.  Smoothed estimates of the posterior moments, $E[\bm{\theta}_{t-k}|D_t]$ and $\text{Cov}[\bm{\theta}_{t-k}|D_t]$, for $k=1, \cdots, t-1$ are trivially obtained. One can also obtain $j$-step-ahead distributions of $(\bm{\theta}_{t+j}|D_{t})$ and  predictive distributions.

{
In the following sections, we illustrate our proposal on both synthetic and real data, highlighting the key components of its analytical development. Section \ref{case} presents details on the multinomial dynamic model, and Section \ref{illust} covers the Normal and Gamma dynamic models.

For an additional illustration, see Section \ref{sales} of the Supplementary Material (available at \if0\blind{\url{github.com/silvaneojunior/paper_kparam}}\else{<Omitted for the sake of anonymity>}\fi) which presents a Poisson application for count data comparing the $k$DGLM with the Conjugate Update of \cite{west1985dynamic} and the local-level approach of \cite{gamerman2013non}. In this univariate setting, $k$DGLM matches the Conjugate Update in recovering the series pattern and in computational time, outperforming the local-level method on both aspects.}

\medskip

\begin{algorithm}[H]
\footnotesize
\textit{\bd{Step 1:} Evolution. Given $\mathbf{m}_{t-1}$, $\mathbf{C}_{t-1}$, and $ \bm{F}_t'\bm{\theta}_t = \bm{\lambda}_t$, compute:}
    \vskip -1.25cm
\begin{eqnarray*}
\boldsymbol{a}_{t} &=& \mathbf{G}_t\mathbf{m}_{t-1};  \; \; \; \;
\boldsymbol{R}_{t} = \boldsymbol{G}_{t} \boldsymbol{C}_{t-1} \boldsymbol{G}_{t}^{\prime}+\boldsymbol{W}_{t} \\
\boldsymbol{f}_{t} &=& \boldsymbol{F}_{t}^{\prime} \boldsymbol{a}_{t}; \; \;\; \;
\boldsymbol{Q}_{t} = \boldsymbol{F}_{t}^{\prime} \boldsymbol{R}_{t} \boldsymbol{F}_{t}. 
\end{eqnarray*}
    \vskip -.5cm
\textit{\bd{Step 2:} Given $\boldsymbol{f}_{t}$ and $\boldsymbol{Q}_{t}$, determine the parameters $\boldsymbol{\tau}_t$ of the conjugate prior:}
\begin{itemize}
    \item \textit{Step 2.1:} Compute the vector of sufficient statistics $\mathbf{H}_q\PC{\bm{\eta}_t}$ for the conjugate distribution $q(\bm{\eta}_t|\boldsymbol{\tau}_t)$. 
    \item \textit{Step 2.2:} Obtain $\bm{\tau}_t$ such that $
E_p(\mathbf{H}_q\PC{\bm{\eta}_t}) = E_q(\mathbf{H}_q\PC{\bm{\eta}_t})$, where the induced distribution is $ p(\bm{\eta}_t|\boldsymbol{f}_{t}, \boldsymbol{Q}_{t})$.
\end{itemize}
\textit{\bd{Step 3:} Update the distribution of $\bm{\eta}_t$ using Bayes' theorem, obtaining:}
$\bm{\tau}^{*}_t = \bm{\tau}_t + \bm{h}(\bm{y}_t).
$\\
\textit{\bd{Step 4:} Given $\boldsymbol{\tau}^{*}_t$, determine the updated parameters $\bm{f}^{*}_{t}$ and $\bm{Q}^{*}_{t}$ for the linear predictor:}
\begin{itemize}
    \item \textit{Step 4.1:}  Assuming that $q(\bm{\lambda}_t|\bm{f}^{*}_{t}, \bm{Q}^{*}_{t})$ is a $k$-variate normal density, the vector of sufficient statistics is $\mathbf{H}_q\PC{\bm{\lambda}_t} = [\bm{\lambda}_t, \bm{\lambda}_t \bm{\lambda}_t']$.
    \item \textit{Step 4.2:} Determine $(\bm{f}^{*}_{t}, \bm{Q}^{*}_{t})$ such that $E_p(\mathbf{H}_q\PC{\bm{\lambda}_t}) = E_q(\mathbf{H}_q\PC{\bm{\lambda}_t})$, where the induced distribution is $ p(\bm{\lambda}_t|\bm{\tau}^*_t) = p_{\bm{\eta}_t|\bm{\tau}^*_t}(\bm{g}(\bm{\eta}_t)) \begin{vmatrix} \nabla_{\bm{\lambda}_t}\bm{g}(\bm{\eta}_t) \end{vmatrix}$.
\end{itemize}
\textit{\bd{Step 5:} Obtain the posterior state moments:}
\vskip -1.25cm
\begin{eqnarray*}
\boldsymbol{m}_{t} &=& \boldsymbol{a}_{t} + \boldsymbol{R}_{t} \boldsymbol{F}_{t} \mathbf{Q}_{t}^{-1} \left(\boldsymbol{f}_{t}^{*} - \boldsymbol{f}_{t}\right); \;\;\;\;
\boldsymbol{C}_{t} = \boldsymbol{R}_{t} + \boldsymbol{R}_{t} \boldsymbol{F}_{t} \mathbf{Q}_{t}^{-1} \left(\boldsymbol{Q}_{t}^{*} - \boldsymbol{Q}_{t}\right) \mathbf{Q}_{t}^{-1} \boldsymbol{F}_{t}^{\prime} \boldsymbol{R}_{t}.
\end{eqnarray*}
\caption{Filtering and one-step ahead prediction in $k$DGLMs}
\label{algo:b_v2}
\end{algorithm}

\medskip 

\begin{algorithm}[H]
{\footnotesize \textit{\textbf{Step 1:} Given $\boldsymbol{a}_{t}$, $\boldsymbol{R}_{t}$, $\boldsymbol{m}_{t}$, and $\boldsymbol{C}_{t}$, compute the smoothed posterior moments for $t=1, \ldots, T$:}
 \vskip -.25cm 
$$
\begin{aligned}
\boldsymbol{m}_{t}^{s} &= \boldsymbol{m}_{t} + \boldsymbol{C}_{t} \boldsymbol{G}_{t+1}^{\prime} \boldsymbol{R}_{t+1}^{-1} \left(\boldsymbol{m}_{t+1}^{s} - \boldsymbol{a}_{t+1}\right), \\
\boldsymbol{C}_{t}^{s} &= \boldsymbol{C}_{t} + \boldsymbol{C}_{t} \boldsymbol{G}_{t+1}^{\prime} \boldsymbol{R}_{t+1}^{-1} \left(\boldsymbol{C}_{t+1}^{s} - \boldsymbol{R}_{t+1}\right) \boldsymbol{R}_{t+1}^{-1} \boldsymbol{G}_{t+1} \boldsymbol{C}_{t}.
\end{aligned}
$$
 \vskip -.25cm
where
$
\begin{aligned}
\boldsymbol{m}_{T}^{s} &= \boldsymbol{m}_{T}, & \boldsymbol{C}_{T}^{s} &= \boldsymbol{C}_{T}.
\end{aligned}
$
\linebreak
{\textit{\textbf{Step 2:} Given $\boldsymbol{m}_{T}$, $\boldsymbol{a}_{T}$, $\boldsymbol{R}_{T}$, $\boldsymbol{C}_{T}$, $\boldsymbol{F}_{t+j}$, and $\mathbf{G}_{t+j}$, compute the $j$-steps-ahead prior for the linear predictor:}
$$
\begin{aligned}
\boldsymbol{f}_{t}(j) &= \boldsymbol{F}_{t+j}^{\prime} \boldsymbol{a}_{t}(j); \;\;\;\;
\mathbf{Q}_{t}(j) = \boldsymbol{F}_{t+j}^{\prime} \boldsymbol{R}_{t}(j) \boldsymbol{F}_{t+j}, \quad j = 1, \ldots, J.
\end{aligned}
$$
where $\boldsymbol{a}_{t}(j) = \mathbf{G}_{t+j} \boldsymbol{a}_{t}(j-1)$ and $\boldsymbol{R}_{t}(j) = \boldsymbol{G}_{t+j} \boldsymbol{R}_{t}(j-1) \mathbf{G}_{t+j}^{\prime} + \boldsymbol{W}_{t+j}$, with $\boldsymbol{a}_{t}(0) = \boldsymbol{m}_{T}$ and $\boldsymbol{R}_{t}(0) = \boldsymbol{C}_{T}$.}}
\caption{Smoothing and $J$-steps-ahead forecast distributions}
\label{algo:b_v3}
\end{algorithm}

\medskip

\section{$k$-Parametric Multivariate Response: Dynamic Multinomial model}\label{case}

This section highlights the $k$-parametric scope of our framework, which distinguishes it from some of the existing sequential, low-cost alternatives. We illustrate the approach with dynamic multinomial models. For a multinomial outcome with $k$ categories, the model entails $k-1$ linear predictors. In our formulation, each predictor can include its own covariates and state evolution, making sequential inference viable for vector time series with rich predictive structure.

In Subsection \ref{calc_multinom} we present details on the analytical development  required for the multinomial particular case of  our proposal, as a template for the general workflow. We have implemented several special cases of the $k$DGLM in the \texttt{kDGLM} package. We present the full derivation of special cases both to elucidate the implemented methodology and to guide users who may wish to extend the approach to other exponential-family distributions not yet supported by the package.

In Subsection \ref{multinom_simul} we validate the $k$DGLM approximation via a simulated study, comparing our posterior summaries to those obtained in Stan \citep[NUTS,][]{STAN2017}. The results show close agreement at substantially lower computational cost.

Finally, Subsection \ref{multinom_applied} applies the method to Brazilian hospital admissions. The example demonstrates the flexibility of the approach, allowing distinct dynamics and covariate effects for each age category of the response.

For notational simplicity, temporal indices are occasionally omitted in what follows; however, it should be noted that all operations are performed at each time $t$.

\subsection{$k$DGLM for Multinomial Models: Analytical Development}\label{calc_multinom}

 The implementation of our sequential approach follows the algorithms outlined in Section \ref{sec:OurProp}. Let $\mathbf{y} | \bm{\pi} \sim \text{Multinomial}(N, \bm{\pi})$,  $\bm{\pi} =  \bm{\eta} = (\pi_1, \pi_2, \ldots, \pi_{d+1})$, where $0 < \pi_l < 1$, $\sum_{l=1}^d \pi_l \leq 1$, $\pi_{d+1} = 1 - \sum_{l=1}^d \pi_l$, $\sum_{l=1}^d y_l \leq N$, and ${y}_{d+1} = N - \sum_{l=1}^d {y}_l$. The conjugate prior is a Dirichlet density: $q(\bm{\pi}|\bm{\tau}) = \frac{1}{\prod_{l=1}^{d+1} \pi_l} \exp \left[\sum_{l=1}^d \tau_l \log \left(\frac{\pi_l}{\pi_{d+1}}\right) + \tau_0 \log \pi_{d+1} + \log B(\bm{\tau}) \right]$, with $B(\bm{\tau}) = \frac{\Gamma(\tau_0)}{\prod_{l=1}^{d+1} \Gamma(\tau_l)}$. Define the vector of linear predictors as $\bm{\lambda} =
 (\lambda_{1}, \ldots, \lambda_{d})' = \PR{\log \left(\frac{\pi_{1}}{\pi_{d+1}}\right), \ldots, \log \left(\frac{\pi_{d}}{\pi_{d+1}}\right)}' = \bm{F}' \bm{\theta}$ and assume that $\bm{\lambda} \sim N_d(\bm{f}, \bm{Q})$.

The sufficient statistics vector under the conjugate prior, used in Algorithm \ref{algo:b_v2},  are:
\[
\bm{H}_q(\bm{\pi}) = \left(\log \left(\frac{\pi_1}{\pi_{d+1}}\right), \ldots, \log \left(\frac{\pi_d}{\pi_{d+1}}\right), \log (\pi_{d+1})\right).
\]
Thus, the system to be solved in Step 2.2 is $ E_q[\bm{H}_q(\bm{\pi})] = E_p[\bm{H}_q(\bm{\pi})] $, where $ p $ is the prior density derived from the normal specification for the linear predictors. This system reduces to:
\[
\begin{cases}
\psi(\tau_i) - \psi \left(\sum_{l=1}^{d+1} \tau_l \right) &= f_i, \quad i = 1, \ldots, d, \\
\psi(\tau_{d+1}) - \psi \left(\sum_{l=1}^{d+1} \tau_l \right) &\approx \log \left(\frac{1}{1 + \sum_{l=1}^d e^{f_l}} \right) + \frac{1}{2} \text{tr}(\tilde{\bm{H}} \bm{Q}),
\end{cases}
\]
where $\tilde{\bm{H}}$ is the Hessian of $ f(\lambda_1,\ldots,\lambda_d)=\log \left(\frac{1}{1 + \sum_{l=1}^d e^{\lambda_l}} \right) $, and $\psi(\cdot)$ is the digamma function, which is approximated by: 
\begin{equation}
\psi(x) \approx \log(x)-\frac{1}{2x}-\frac{1}{12x^2}.\label{digammaprox}
\end{equation}

Step 3 updates the conjugate prior hyperparameters to $\tau_l^* = \tau_l + {y}_l$ for $l = 1, \ldots, d$ and $\tau_0^* = \tau_0 + N$. Now, let $q$ be a multivariate normal density for the vector of linear predictors, with sufficient statistics vector 
$\bm{H}_q  = (\bm{\lambda}, \bm{\lambda \lambda'})'$. In Step 4.2, the system $ E_q[\bm{H}_q(\bm{\lambda})] = E_p[\bm{H}_q(\bm{\lambda})] $ must be solved, where, now, $p$ is the updated conjugate distribution. This reduces to: $
f_l^* = \psi(\tau_l^*) - \psi(\tau_{d+1}^*) \,\mbox{and}  \,
Q^*_{ll'} = \psi'(\tau_l^*) I_{(l=l')} + \psi'(\tau_{d+1}^*),$
where $\psi'(\cdot)$ denotes the trigamma function and $I_{(l=l')}$ is the indicator function of $l=l'$.

Finally, the updated moments of the states $\bm{\theta}$ are obtained assuming a joint normal distribution for $\bm{\lambda}$ and $\bm{\theta}$ and using properties of the normal distribution to obtain the conditional distribution of $\bm{\theta}$ on $\bm{\lambda}$ \citep[see][p. 639]{west1997}.  Predictions of future realizations of the observables are trivially obtained using the predictive  Dirichlet-multinomial($N$,$\bm{\tau}$) density. Sequential updating for binary classification time series (a distinguished particular case of the multinomial structure) is detailed in Section \ref{Bernoulli} of the supplementary material (\if0\blind{\url{github.com/silvaneojunior/paper_kparam}}\else{<Omitted for the sake of anonymity>}\fi).

 \subsection{Simulated Study: Dynamic Multinomial Model}\label{multinom_simul}

The efficiency of the method was evaluated through the generation of synthetic data from the model:
 \begin{equation} 
 \begin{aligned} {y}_1,{y}_2,{y}_3|\bm{\eta}_{t} &\sim \text{Multinomial}(N,\bm{\eta}_t), \\ \log\left(\frac{\eta_{i,t}}{\eta_{3,t}}\right) &= \theta_{i,t}, i=1,2,\\ \theta_{i,t} &= \theta_{i,t} + \omega_{i,t}, \forall i, \end{aligned} \end{equation} with $\omega_{1,t},\omega_{2,t} \overset{\mathrm{iid}}{\sim} \mathcal{N}(0,W_t),$ and $W_t=0.01$. The number of categorical allocations for the response was set at $ N = 1, 2, 5, 10, 20, 50, 100 $. The frequentist properties of the inference produced by the proposed sequential approximation  were evaluated by generating 500
 data replicates for each $ N $. Both our sequential approach and non sequential stochastic simulation via NUTS \citep{Hoffman2014} in Stan \citep{STAN2017} were used to fit the model for each sample replicate.

\begin{figure}[htb!]
    \centering
    \includegraphics[width = 1\linewidth]{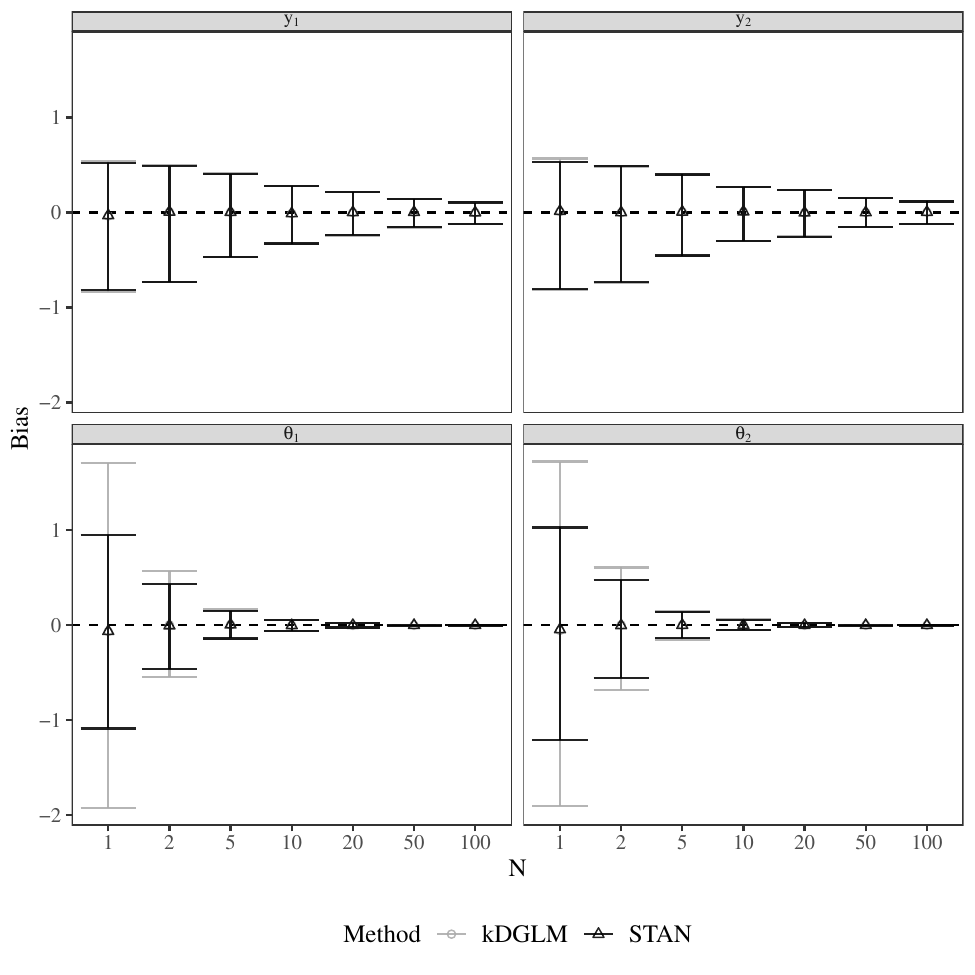}
    \caption{The bias of the one-step-ahead prediction of ${y}_{1,T+1}$ and ${y}_{2,T+1}$, for $T=100$. Triangles represent the average bias across 500 sample replicates, while the error bars represent the $0.975$ and $0.025$ quantiles for the bias across the replicates. To facilitate visualization, the bias is  divided by $N$. Gray: Sequential kDGLM. Black: NUTS via Stan.}
    \label{fig:FigSimulMult_texto}
\end{figure}

Figure \ref{fig:FigSimulMult_texto} presents a frequentist analysis of the bias in one-step-ahead predictions of $\mathbf{y}_{1,T+1}$ and $\mathbf{y}_{2,T+1}$, and in smoothed estimates of $\theta_{1,T}$ and $\theta_{2,T}$, for $T=100$.
  Additional analyses for different sample sizes $T$, varying from $T=2$ to $T=50$, are presented in Section \ref{Mult_simul} of the supplementary material.  The results show that the sample size $T$ has little impact on the approximation quality of our method. As for the number of categorical allocations $ N $, the range of the bias across replicates approximates the ranges obtained via NUTS, as $ N $ increases. As observed, for $ N \geq 5 $, the inference obtained through the proposed method is approximately equivalent to that derived using  NUTS, for sample sizes varying from $T=2$ to $T=100$. 

  In Section \ref{case}, a  multinomial dynamic model with trend and seasonal components is applied to real data in an epidemiological context to demonstrate the practical utility of the dynamic multinomial formulation, as well as the flexibility of the proposed method in accommodating structures with varying dynamic structural blocks for predictors in $k$-parametric models.

\subsection{Application: Hospital Admissions in Brazil}\label{multinom_applied}
Acute diarrheal disease, exacerbated by poor living conditions and malnutrition, poses a significant health threat in Brazil. Current monitoring focuses on non-cholera bacterial, viral, and parasitic outbreaks spread through food, water, and interpersonal contact. Dynamic Bayesian models facilitate monitoring epidemiological trends, as well as accommodating structural shifts, predicting outbreaks, and quantifying intervention impacts. 

Rotavirus is a major etiological agent of diarrheal  disease, causing a substantial proportion of severe cases, hospitalizations, and deaths in Brazil \citep{ambrosini2012}. Since March 2006, the Rotarix$^@$ vaccine has been administered to children, significantly reducing hospitalizations and deaths in vaccinated age groups \citep{Linhares2013}. \citet{ambrosini2012} reported substantial reductions in diarrhea-related hospitalization rates in children up to age five after vaccination.

We analyze monthly hospital admission data for diarrhea and gastroenteritis, in Brazil, from 2000 to 2022, thus including the COVID-19 pandemic period, using a dynamic multinomial model. This model captures age-specific patterns and assesses the contributions of different age groups to the pool of hospital admissions, particularly infants (<1 year) and young children (1-4 years), the age groups targeted by the rotavirus vaccine. The data are available at http://tabnet.datasus.gov.br.

A preliminary cluster analysis of hospital admissions due to diarrhea and gastroenteritis was applied, to detect correlations among time series across different age groups. Our main focus is on hospital admissions for infants, and the cluster analysis resulted in the following age groups: infants ($<$1 year); young children (1-4 years); children (5-14 years); youths/adults (15-59 years - reference group); and $\ge$60 years (seniors). Time series for the first two groups are exhibited as solid circles in Figure \ref{both_groups_one_step}. 

A dynamic multinomial model was fitted to the categorized data. Let ${\mathbf{y}}_t=({y}_{1t},\ldots, {y}_{5t})$, where ${y}_{jt}$ denotes the number of hospital admissions for the age group $j$, $j=1, 2, \ldots, 5.$     Assume that ${\mathbf{y}}_t|\eta_{1t}, \eta_{2t}, \eta_{3t}, \eta_{4t}, \eta_{5t}\sim \text{Multinomial}(\eta_{jt},\,\, j=1:5)$ where $\eta_{5t} = \PC{1 - \sum_{j=1}^4 \eta_{jt}}$ with predictive structure:$\lambda_{jt}   = \log\PC{\frac{\eta_{jt}}{\eta_{5t}}}  = F_{jt}'\theta_{jt}, \; j=1,2,3,4. $ The same dynamic structure, which was considered flexible enough to capture observed  patterns in the real data, was adopted for each category: $F'_{jt} = \PR{1,0,1,0,1}$;  $G_{jt} =  {blockdiag}\PR{G_{trend}, G_{seasonal},G_{noise}}, \; j=1,\ldots,4$, where:
$
G_{trend} = \left[\begin{array}{cc} 
1 & 1 \\
0 & 1 \\
\end{array}\right];\;  G_{seasonal} = \left[\begin{array}{cc} 
 \cos(\omega) & sin(\omega)\\
 -sin(\omega) & \cos(\omega)\\
\end{array}\right] , \; \omega = \frac{2\pi}{12},\; G_{noise} = 0.$ Therefore, $F'_t = blockdiag\PR{F'_{1t}, F'_{2t}, F'_{3t}, F'_{4t}}$ and $G_t = blockdiag\PR{G_{1t}, G_{2t}, G_{3t}, G_{4t}}$. Discount factors 0.95 and 0.975 were assigned for trend and seasonality, respectively. A noise component with variance $0.005$ was included to accommodate overdispersion.

\if0\nofig{
\begin{figure}[!htb]
\centering
\includegraphics[width = 1\linewidth]{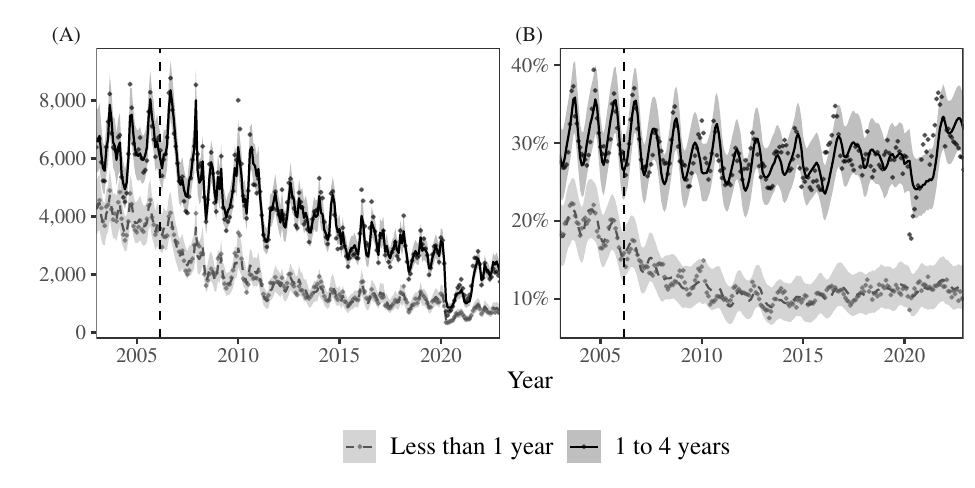}
    \caption{Observed values (solid circles)  compared with one-step-ahead predictions for age groups 1 and 2: number of admissions  (A); probability of hospital admission for each age group (B). The vertical dashed line indicates the moment of vaccine introduction.}
 \label{both_groups_one_step}
\end{figure}
}\else \begin{figure}\caption{}\label{both_groups_one_step}\end{figure}
\fi

The multinomial dynamic model provided estimates of compositional probabilities for different age groups, conditional on hospitalization due to diarrheal disease or gastroenteritis. Sequential updating was carried out following the proposed kDGLM procedure detailed in Section \ref{calc_multinom}. Using data from the Brazilian Institute of Geography and Statistics (IBGE) and Datasus, we estimated hospital admission probabilities conditionally on each age group (see Section \ref{Offset} in the supplementary material for details).

As can be seen in Figure \ref{both_groups_one_step}, our findings are consistent with the Brazilian epidemiological literature, showing a marked decline in hospital admission counts (Panel A) and probabilities (Panel B) for infants following the introduction of the rotavirus vaccine in the National Immunisation Plan. A similar decline is observed in admission counts and probabilities for the 1–4 year age group. Interactive graphs presenting predictive results for all age groups are  available at \if0\blind{\url{github.com/silvaneojunior/paper_kparam}}\else{<Omitted for the sake of anonymity>}\fi. After vaccine introduction, a reduction in the probability of hospitalization due to diarrhea and gastroenteritis is evident across all age groups, with immediate effects for infants and gradual declines for older groups. This scaling effect likely reflects reduced rotavirus transmission and the progression of immunized individuals into subsequent age groups. The model also accounts for structural changes, such as the Covid-19 pandemic, capturing a sharp decline in hospital admissions during lockdown and adapting to evolving patterns as restrictions were eased.

\section{$k$-Parametric Univariate Responses: Dynamic Normal and Gamma Models}\label{illust}

This section shows how the proposed framework works for {univariate} $k$-parametric DGLMs while accommodating rich dynamics (e.g., trend and seasonality) and allowing multiple observational parameters—such as mean and precision—to evolve over time. The distribution-specific algebra required by Algorithms~\ref{algo:b_v2}--\ref{algo:b_v3} was implemented in the \texttt{kDGLM} R package \citep{santos2024}.

We focus on stochastic volatility (SV) settings and present two alternative formulations: a {normal} model that treats returns $y_t$ on their original scale with dynamic mean and log-precision, and a {gamma} model for squared returns. The normal formulation preserves return signs, whereas the gamma specification connects directly to the classical SV representation via $\log(y_t^2)$. Section \ref{Proof} in the supplementary material (\if0\blind{\url{github.com/silvaneojunior/paper_kparam}}\else{<Omitted for the sake of anonymity>}\fi) provides an in-depth examination of the bi-parametric normal particular case of our proposal.

We assess accuracy and efficiency in a controlled simulation with known latent states (Section~\ref{artificial}), comparing our sequential updates to NUTS in Stan \cite{Hoffman2014, STAN2017}, a particle filter \cite{Liu2016KernelEnGMF}, Variational Bayes \citep{Kucukelbir2015}, and INLA \citep{Rue2009INLA,Rue2017}. Finally, we apply gamma and normal   formulations to monthly IBM returns (Section~\ref{real}). Across these illustrations, the sequential approach delivers posterior summaries comparable to strong baselines while maintaining linear-time updates and low memory use,  extending classical uni-parametric approaches (e.g., \citealp{west1985dynamic}) to the broader class of $k$-parametric exponential-family models.

\subsection{Stochastic Volatility Models}

Financial decision making often relies on volatility as a risk indicator. The foundations of SV models are strongly related to financial theories, and from a practical point of view, they are able to appropriately capture the main empirical properties of financial return series. A traditional return model is ${y}_t=e^{h_t/2}\epsilon_t$, $\epsilon_t \sim N(0,1)$, or $\log(y^2_t) = h_t + a_t$, where $h_t$ follows an AR(1) process with autoregressive coefficient $\gamma$ and $a_t$ follows a log-$\chi^2$ distribution,  usually approximated by a mixture of normals \citep{chib2009mult}. We propose two dynamic approaches within the $k$-parametric exponential family. The first employs a normal model, considering the returns at their original scale as the observables, associating their mean and precision at each time $t$ with dynamic structures. The second approach applies a dynamic gamma model for the squared returns.

\clearpage

\centerline{\bf Normal biparametric dynamic model for the returns}\label{volat_normal}

\medskip

The implementation of our approach for univariate Gaussian responses, with dynamic predictive structure for the mean and log-precision, follows Algorithm \ref{algo:b_v2}  described in Section \ref{sec:OurProp}. The bidimensional parameter vector is denoted by
$\bm{\eta}=(\mu,\phi)$ and  $(y|\bm{\eta}) \sim N(\mu,\phi^{-1}), \, \mu \in \Re,\, \phi>0$. The conjugate prior is a normal-gamma density, hierarchically structured as: $\mu|\phi \sim N [\mu_0, (c_0\phi)^{-1}], \, \, \, \phi \sim {\cal G}(n_0/2, d_0/2).$
Consider a vector of linear predictors: $\bm{\lambda}(\bm{\eta})=( \lambda_{1}, \lambda_{2})'= (\mu,  \log(\phi) )'=(F_{1}'\theta_{1},F_{2}'\theta_{2})'$
and the reparameterization:
$    c_0=-2\tau_1, \quad 
    \mu_0=-\frac{\tau_2}{2\tau_1}, \quad
    \frac{d_0}{2}= \frac{\tau_2^2}{4\tau_1}-\tau_3 \quad \mbox{and} \quad
    \frac{n_0}{2}= \tau_0+1/2.
  $
We follow Algorithm \ref{algo:b_v2}, noting that the conjugate prior sufficient statistic vector is  ${\bm{H}}_q(\bm{\eta})=(\phi\mu^2,\phi\mu,\phi,\log\phi)'$. Then, the solution of the system in Step 2.2,  $E_q[\bm{H}_q(\bm{\eta})]=E_p[\bm{H}_q(\bm{\eta})]$, is:
\begin{eqnarray*}
    \left\{ \begin{array}{lll}
    \tau_0&=&\frac{1}{q_2}-\frac{1}{2},  \,\,\,\,\,\,\,\,\,\,\,\,\,\,\,\,\,\,\,\,\,\,\,\,\,
     \,\, \, \tau_1=-[2q_1\exp(f_2+Q_2/2)]^{-1}  \\
     \tau_2&=-&2\tau_1 (f_1+Q_{12}), \,\,\,\,\,\,\,\,\,\,\tau_3=(f_1+q_{12})\tau_2 - [q_2\exp(f_2+Q_2/2)]^{-1}.
      \end{array}\right.
      \label{eqs_ida_normal_final}
    \end{eqnarray*}
    Once a new observation $y$ is available, the conjugate prior hyperparameters are updated to $\bm{\tau}^*=(\tau_1^*,\tau_2^*,\tau_3^*, \tau_0^*)=(\tau_1-1/2,\tau_2+y,\tau_3-y^2/2,\tau_0+1/2)$. At this stage,  one must  reconcile the posterior distribution for the linear predictors implied by the conjugate distribution for $\bm{\eta}$ with a multivariate normal.  
Let $q$ denote a multivariate  normal density for the  vector of linear predictors, with sufficient statistic vector: $\bm{H}_q'=(\bm{\lambda},\bm{\lambda\lambda'})$. Step 4.2 reduces to solving the system $E_q[\bm{H}_q(\bm{\lambda})]=E_p[\bm{H}_q(\bm{\lambda})]$, where $p$ is the updated distribution, resulting in:
\begin{equation*}
    \left\{ \begin{array}{lll}
    f_1^*&=& \mu^*_0  \,\,\,\,\,\,\,\,\,\,\,\,\,\,\,\,\,\,
    f_2^*  = \psi(n^*_0/2)-\log\PC{d^*_0/2}\nonumber \nonumber \\
     Q_{1}^*&=&\frac{d_0^*}{c^*_0n^*_0} \,\,\,\,\,\,\,\,\,\,\,\,\,Q_{12}^*=0\,\,\,\,\,\,\,\,\,\,\,\,\ Q_{2}^*=\psi'(n_0^*/2).
 \end{array}\right.
    \end{equation*}  

The updated moments of the  states $\bm{\theta}$ are trivially obtained by applying normal distribution properties. The predictive distribution for ${y}_t$ is a Student-t density with $n_0$ degrees of freedom, location parameter $\mu_0$ and scale parameter $d_0/c_0n_0$, which can be evaluated when step 2.2 is completed. In order to guarantee that the system in Step 4.2 has a solution, we imposed a restriction on the parameter space. Details can be found in the supplementary material, Section \ref{Proof}.

For the specific purpose of modeling returns, let ${y}_t \sim N(\mu_t, \phi_t^{-1})$ and assume the following predictive dynamic structure for the mean returns $\mu_t$  and their log-precision $\log(\phi_t)=-h_t$: $\bm{\lambda}_t =  ( \lambda_{1t}, \lambda_{2t}  )'  = ( \mu_t, \,  -h_t  )'; \; \;
 \mbox{with evolution }( \mu_{t} ,  h_{t}, \gamma_t   )'= ( \mu_{t-1},  \gamma_{t-1} h_{t-1},  \gamma_{t-1} )' +(\omega_{1t}, \,\omega_{2t}, \,
\omega_{3t})',$
where $\gamma_t=\gamma$ is an autoregressive coefficient estimated according to the ideas in \citet[][Chapter 13]{west1997} which are naturally accommodated in our approach assuming $\omega_{3t}=0$ with probability 1.    In the applications exhibited in Subsections \ref{artificial} and \ref{real}, we adopted a normal prior for $\tanh^{-1}\{\gamma\}$, constraining $\gamma$ to the interval (-1,1), so that  $h_t$ is a stationary process. 

\bigskip

\centerline{\bf  Gamma dynamic model for squared returns}

\medskip

Considering the traditional SV model,   $\log (y^2_t) = h_t + a_t$, with $a_t \sim log-\chi^2$, it follows that ${y}_t^2 \sim \mathcal{G}\left(\frac{1}{2},\frac{1}{2}\exp(-h_t)\right)$. Thus, the particular gamma case of our approach  can be applied  to model ${y}_t^2$, with shape parameter  $\alpha_t=\alpha=\frac{1}{2}$ and $\mu_t=E[{y}_t^2]=e^{h_t}$. Then ${\lambda}_t   =   \log(\mu_t)    =  h_t$, with evolution given by  $h_t=\gamma h_{t-1} +\omega_t$, where $\gamma$ is an autoregressive parameter estimated following \citet[][Chapter 13]{west1997}.

Gamma dynamic models with known shape parameter $\alpha$  were implemented in \texttt{kDGLM} package 
 by following Algorithm \ref{algo:b_v2} described in Section \ref{sec:OurProp}.  The  conjugate prior for {$\eta=\mu$} is an inverse-gamma density $q(\eta|\bm{\tau})= \frac{1}{\eta} \exp[-\tau_0 \log(\eta) - \tau_1/\eta - b(\tau_0, \tau_1)]$, where  $b(\tau_0, \tau_1)=\log(\Gamma(\tau_0))  -\tau_0 \log(\tau_1)$. Consider the linear predictor: $\lambda=\log(\mu)= \bm{F}'\bm{\theta}$ and the conjugate prior sufficient statistic vector: ${\bm{H}}'_q(\eta)=(-\log(\eta),-1/\eta)$.
Following Algorithm \ref{algo:b_v2} and using the fact that $E_q[\bm{H}_q(\eta)]=\nabla b(\tau_0,\tau_1)$ as well as approximation (\ref{digammaprox}) for the digamma function, the solution of the system $E_q[\bm{H}_q(\eta)]=E_p[\bm{H}_q(\eta)]$ in Step 2.2 results in $
 \tau_1 =   \tau_0\exp(f-Q/2)\,\,\, \mbox{and} \,\,\,  \tau_0 = (1+\sqrt{1+2Q/3})/(2Q)
$.  The conjugate prior hyperparameters are updated to  
$\tau_1^*=\tau_1+y$ and $\tau_0^*=\tau_0+\alpha$. Step 4.2 is completed by solving the system $E_q[\bm{H}_q(\bm{\lambda})]=E_p[\bm{H}_q(\bm{\lambda})]$, where, now,  $p$ is the updated conjugate  distribution and $q$ denotes a normal density for the  linear predictor, with sufficient statistic vector: $\bm{H}_q'=({\lambda},{\lambda^2})$. The following systems are generated: $f^*=\log(\tau_1^*) -\psi(\tau_0^*)$ and $ Q^*=\psi'(\tau_0^*)$.

The predictive density for ${y}_t$ is an inverse beta distribution with shape parameters $\tau_0$ and $\alpha$ and scale parameter $\tau_1$, which can be evaluated once step 2.2 is completed. Note that from a probabilistic standpoint, the gamma and normal specifications for the volatility are equivalent, since if ${y}_t \sim N(0, e^{h_t})$, it follows that ${y}_t^2 \sim {\cal G}\left( {1}/{2}, {e^{-h_t}}/{2} \right) $, but, as previously mentioned, the normal formulation preserves the signals of the returns. 
 
In Subsection \ref{artificial}, the proposed models and inferential approach are applied to artificial data, aiming to evaluate the performance of the normal and gamma dynamic formulations in capturing the volatility structure, as well as to compare the performance of our proposal to the results obtained via NUTS in Stan \citep{STAN2017}. 
The gamma and normal formulations based on our sequential  method are then applied to a real dataset, as shown in Subsection \ref{real}.

\subsection{Simulated Study: Normal and Gamma Models}\label{artificial}

We simulated a time series of 365 observations, based on a dynamic normal structure with $\gamma = 0.95$, in order to evaluate the effectiveness of our method in recovering latent components of a stochastic volatility model. In order to perform a direct comparison of our approach with NUTS via Stan using  identical models, we applied the normal formulation with known evolution variance $\text{Var}(\omega_{2t})=0.05$   for the state $h_t$.

As  shown in panel A of Figure \ref{fig:compara_simul}, the volatility estimations under our sequential updating method for the normal and gamma models were nearly identical, as anticipated due to their equivalence. The proposed approach effectively captures volatility fluctuations, although a few time periods show exceptions.  Panel B exhibits the comparison of our sequential updating method for the normal dynamic model with non-sequential NUTS via Stan. Panel C displays a comparison between the proposed method and non-sequential variational inference implemented in Stan.  Posterior summaries for the autoregressive coefficient $\gamma$ are presented in Table \ref{tabvolat1}. The sequential updates of the normal and gamma models produced nearly identical estimates, close to those for the NUTS fit of the normal formulation, which is taken as a golden standard in this analysis.  The true value $\gamma = 0.95$ was covered by the $95\%$ credibility intervals across all scenarios.  
 Stan required 26.131 seconds - more than 150 times longer than our method (0.151 seconds). The computational time of our approach scales linearly with the length of the time series, ensuring efficiency for long series. All computational times reported in this paper were measured with an i7-9750H CPU with 24 GB of RAM.

A particle filter was implemented following\cite{liu_west_2001} 
with 50{,}000 particles, to ensure adequate representation of the posterior distribution. The procedure required 12.07 seconds -- over thirty times longer than the proposed kDGLM method. As a sequential algorithm, the particle filter allows marginal cost assessment, with an average update time of approximately 0.033 seconds per new observation. Despite this advantage, particle filters are sensitive to abrupt structural changes, require careful tuning (e.g., the number of particles, which strongly affects computational cost), and typically perform poorly in the presence of static components. By contrast, the proposed method does not require Monte Carlo sampling or tuning, and yields closed-form approximations that are analytically tractable and numerically stable. It handles both static components and sudden dynamic shifts efficiently, ensuring scalability to large datasets.
Variational Bayes (VB) using a mean-field approximation required 4.25 seconds -- almost thirty times slower than the proposed method. Finally, we compared the proposed method to INLA-based alternatives. INLA’s core engine performs analytic or semi-analytic Laplace approximations for latent-Gaussian models with exponential-family likelihoods. Dispersion parameters, when present, are treated as global hyperparameters and therefore cannot vary over time. The proposed stochastic volatility models do not satisfy these assumptions. To enable comparison, two INLA-based alternatives were fitted: a seven-component Gaussian mixture (Chib’s model), which required 1.06 seconds, and a Gamma model for $y^2$ with prior specification $\theta_1 \sim N(0,\, W/(1-\gamma))$, which required 0.72 seconds. Although these models differ from those implemented in \texttt{kDGLM}, they provide reasonable approximations.  Overall, the proposed method allowed model flexibility and achieved accuracy at a fraction of the computational cost.  

 {

 \if0\nofig{
\begin{figure}[!htb]
\centering
\includegraphics[width = 1.065\linewidth]{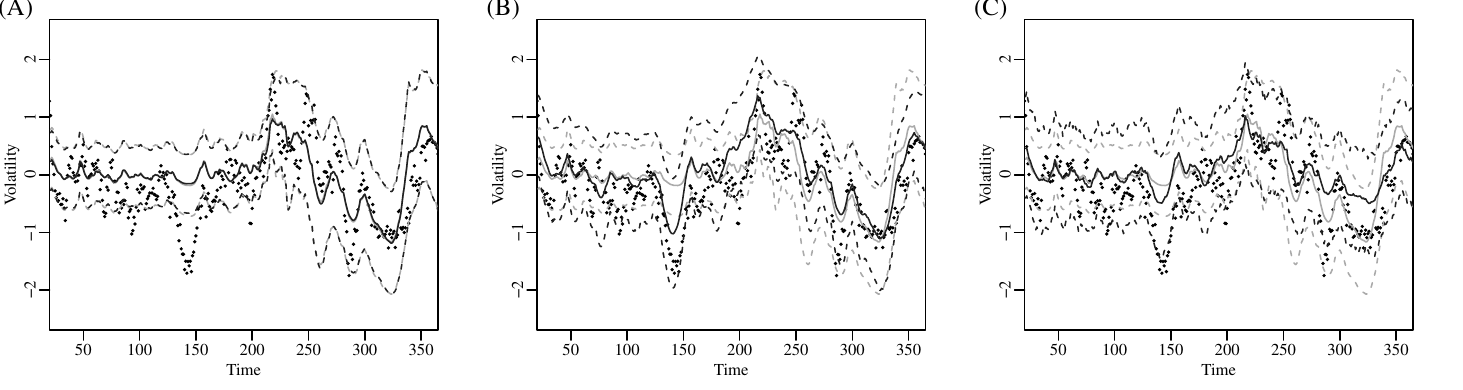}
    \caption{
Comparison of smoothed means (solid lines) and credible intervals (dashed lines) for volatility estimation from synthetic data. The solid circles indicate theoretical volatilities. (A): results of the proposed sequential approach for normal (gray) and gamma (black) formulations. (B): comparison of normal formulation using the sequential approach (gray) and NUTS via Stan (black). (C): comparison of normal formulation using the sequential approach (gray) and Variation Bayes via Stan (black).}
    \label{fig:compara_simul}
\end{figure}
}\else 
\begin{figure}\caption{}\label{fig:compara_simul}\end{figure}
\fi

\if0\nofig{
\begin{table}[!htb]
\centering \small
\begin{tabular}{@{}lrrrr@{}}
\toprule
Model/ Inferential approach & Mean & $2.5\%$ quantile  & $97.5\%$ quantile & Comp. Time \\    \hline   
Gamma/  kDGLM  & 0.937 & 0.900 & 0.974 & 0.105 s\\
Normal/  kDGLM & 0.939 & 0.903 & 0.975 & 0.151 s\\
Normal/ Particle Filter & 0.937 & 0.888 & 0.978 & 2.070 s \\
Normal/ NUTS & 0.937 & 0.885 & 0.978 & 26.131 s\\
Normal/ VB & 0.827 & 0.725 & 0.898 & 4.250 s\\
Gaussian Mix/ INLA & 0.951 & 0.933 & 0.966 & 1.060
s\\
Gamma/ INLA & 0.926 & 0.833 & 0.977 & 0.720 s\\
\bottomrule
\end{tabular}
\caption{Posterior summaries for the autoregressive coefficient ($\gamma$) in the normal and gamma stochastic volatility models for synthetic data, via sequential updating and NUTS via Stan.}
\label{tabvolat1}
\end{table}
}\else \begin{table}\caption{}\label{tabvolat1}\end{table}
\fi

 { For this specific model, the sequential implementation of the kDGLM method employs a linear approximation due to the unknown autoregressive coefficient 
$\gamma$
 in the evolution matrix 
$\mathbf{G}$. Section \ref{S_Stan} of the supplementary material shows that this approximation largely accounts for the discrepancies observed between the results of the proposed method and those obtained using NUTS in Stan, with nearly identical results from both methods if $\gamma$ is fixed at its true value (see Figure \ref{fig:compara_simul_supl}). Thus, for models where the evolution matrix 
$\mathbf{G}$
 is known, the kDGLM method is expected to produce inferences nearly identical to those generated by Markov Chain Monte Carlo schemes.} On the other hand, for long time series analysed using models where 
$\mathbf{G}$ depends on hyper parameters that introduce non linearities in the system, and in contexts demanding real-time inference, a slight loss in accuracy or precision can be outweighed by the substantial computational efficiency offered by the kDGLM method}.

\subsection{Application: IBM Returns}\label{real}

This section presents the results of our proposed method, using normal and gamma dynamic formulations, to fit and forecast monthly IBM returns from January 1926 to December 1999. The data were obtained from \citet{tsay2005analysis}.

The posterior summaries of the autoregressive coefficient $\gamma$ are presented in Table \ref{table_volat_ibm}. Similar to the synthetic data results, the normal and gamma formulations using the proposed sequential approach are nearly indistinguishable. These posterior summaries align with those obtained via NUTS in Stan.

\if0\nofig{
\begin{table}[!htb]
\centering{\small
\begin{tabular}{@{}lrrr@{}}
\toprule
Model/ Inferential approach & Mean & $2.5\%$ quantile  & $97.5\%$ quantile \\ 
\hline
Gamma/  kDGLM  & 0.934 & 0.916 & 0.951\\
Normal/  kDGLM & 0.933 & 0.915 & 0.951\\
Normal/  Particle Filter & 0.938 & 0.910 & 0.969\\
Normal/  NUTS        & 0.938 & 0.906 & 0.965\\
Normal/ VB & 0.834 & 0.796 & 0.867\\
Gaussian Mix/ INLA & 0.972 & 0.964 & 0.979\\
Gamma/ INLA & 0.989 & 0.976 & 0.997
\\
\bottomrule
\end{tabular}
\caption{Posterior summaries of the autoregressive coefficient ($\gamma$) in the normal and gamma stochastic volatility models for the IBM return data, via sequential updating and Hamiltonian Monte Carlo - Stan.}
\label{table_volat_ibm}}
\end{table}
}\else \begin{table}

\label{tabvolat2}\end{table}
\fi

Figure \ref{compara_ibma} displays the smoothed estimates for volatility $ h_t $. The normal and gamma formulations yield almost identical point estimates and credible intervals, as shown in panel A. To objectively compare models, we calculated a Bayes factor and the normal model was superior to the gamma one. The predictive log-likelihood difference between models favored the normal model by $316.3$. As shown in panel B, the sequential estimates for volatilities using both our sequential updating scheme and Stan are consistent, indicating that the proposed method provides adequate approximations with significantly reduced computational time compared to a ``gold standard'' method that ensures convergence to the true posterior. Note that sequential updating via $k$DGLM is trivial, whereas with non-sequential methods such as NUTS, each new observation requires reprocessing the entire time series, resulting in a high computational cost for updating.  Moreover, the sequential formulation of $k$DGLM allows for the incorporation of external information at any time, which can be advantageous in econometric applications.

\if0\nofig{
\begin{figure}[!htb]
\centering
\includegraphics[width = 1.065\linewidth]{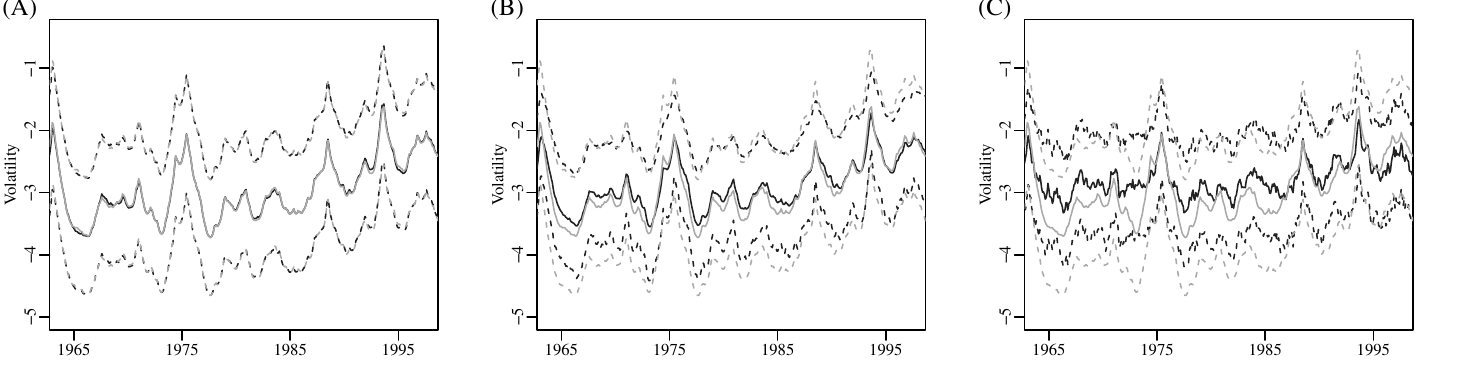}
\caption{Smoothed mean and credibility intervals for IBM volatilities $h_t$. (A): Comparison between normal (grey) and gamma formulations using sequential kDGLM (black). (B): Sequential kDGLM estimates (grey) compared to Hamiltonian Monte Carlo estimates obtained via Stan (black). (C): Sequential kDGLM estimates (grey) compared to Variational Bayes estimates obtained via Stan (black).}\label{compara_ibma}
\end{figure}
}\else \begin{figure}\caption{}\label{compara_ibma}\end{figure}
\fi

 A similar application was proposed by \citet{souza2016extended}, who used the generalized method of moments (GMM) to handle a system with more equations than parameters, but this introduced additional approximations and processing overhead. Our approach avoids these issues.

\section{Concluding Remarks and Future Research}\label{concl}

In this study, we introduced a method for sequentially updating information in {\it dynamic generalized linear models} for  univariate and multivariate responses within the $k$-parametric exponential family. The proposed method builds on the conjugation properties of the exponential family and the Projection Theorem, reconciling conjugate prior and posterior distributions for canonical parameters with distributions induced by normality assumptions for the states in linear predictors. These predictors can accommodate stochastic trends, seasonality, and dynamic covariate effects. Note that our proposal allows the analyst to assign dynamic predictors for all the $k \ge 1$ parameters that characterize an exponential family distribution. This was illustrated, for instance, through dynamic structures associated to both the mean and log-precision of Gaussian responses.  

Compared with alternative approaches, the proposed sequential method achieved results virtually identical to those obtained with non-sequential Bayesian procedures, such as NUTS in Stan, while maintaining  consistency in posterior estimates and credible intervals. Unlike Monte Carlo–based methods and particle filters, which require tuning and may struggle with abrupt structural changes, the proposed algorithm delivers analytically tractable and numerically stable closed-form updates, allowing for key aspects in time series analysis, such as sequential monitoring and intervention. It also compared favorably to mean-field Variational Bayes and INLA, imposing no restrictive assumptions on model structure or distributional form. Overall, the method combines accuracy, robustness, and scalability, providing an efficient and generalizable framework for sequential Bayesian inference in dynamic models. Altogether, these properties make the proposed approach a competitive option for real-time analysis, producing filtered and smoothed distributions as well as forecasts in efficient computational time,  resulting in timely decision making.

The analytical development of the method as well as applied illustrations were presented for specific cases:  multinomial, Bernoulli,  Poisson, gamma, and normal with dynamic predictors for mean and log-precision.

We are currently expanding the proposed method to deal with other observational distributions, including Dirichlet compositional responses and multivariate normal outputs with focus on multivariate stochastic volatility models. The \texttt{kDGLM} R package has been developed to implement the described models, providing users with a tool for efficient dynamic modeling of both univariate and multivariate responses. \texttt{kDGLM} is available in CRAN-R and features automatic monitoring, intervention, autoregressive components, and transfer functions.

\if0\blind
{
  \section*{Acknowledgements}

We thank Professor Heudson Mirandola (UFRJ), for his fruitful collaboration and suggestions, and Professor Alexandra Schmidt (McGill University) for her careful review of a preliminary version of this manuscript and valuable comments.
}\else{
 \section*{Acknowledgements}
<Omitted for the sake of anonymity>
} \fi

\if0\blind
{
  \section*{Funding}
{The research presented in this paper was partially supported by PAPD/UERJ, grant E-26/007/10667/2019/22 and  Funda\c c\~ao de Amparo \`a Pesquisa do Rio de Janeiro (Faperj) grant:  E-26/203.944/2022 - Emeritus Visiting Researcher Fellowship (H. S. Migon), and a CAPES grant (S.V. dos Santos Jr. and R. Marotta).} 
}\else{
\section*{Funding}
{<Omitted for the sake of anonymity>} 
} \fi

\section*{ Declaration of generative AI and AI-assisted technologies in the manuscript preparation process}

During the preparation of this work the authors used the ChatGPT AI tool in order to review and refine the formal written English and to enhance textual conciseness. After using this tool, the authors reviewed and edited the content as needed and take full responsibility for the content of the published article.

\appendix
\renewcommand{\thesection}{\Alph{section}}
\renewcommand{\thetable}{\Alph{section}.\arabic{table}}
\renewcommand{\thefigure}{\Alph{section}.\arabic{figure}}
\setcounter{equation}{0}
\renewcommand{\theequation}{\Alph{section}.\arabic{equation}}

\setcounter{section}{0}

\setcounter{section}{19}

\subsection{ A Poisson Model for Quarterly Sales}\label{sales} 

{This section presents a model for univariate, uniparametric responses. Unlike methods based on local levels, where only the level evolves dynamically, our goal is to emphasize the versatility of the proposed method, which allows the assignment of temporal dynamics for all latent components of the predictor. We highlight the lower computational cost involved in our proposal, in comparison with a local level method \citep{gamerman2013non}. Finally, we show that, in the single parameter case, our method is equivalent to the one suggested by \citet{west1985dynamic}, in terms of model flexibility,  computational cost and resulting inference.}
  
  For the analytical development of our proposed approach for Poisson responses, we follow the steps described in Section \ref{sec:OurProp}. Let $y|\eta \sim Po(\eta), \, \eta>0$. The conjugate prior is a gamma density $q(\eta|\bm{\tau}) = \frac{1}{\eta} \exp[\tau_1 \log(\eta) - \tau_0  \eta - b(\tau_0, \tau_1)]$, where $b(\tau_0, \tau_1)= \log(\Gamma(\tau_1))  -\tau_1 \log(\tau_0)$. Consider the linear predictor $\lambda= \log(\eta)=\bm{F}'\bm{\theta}$ and the conjugate prior sufficient statistics vector: ${\bm{H}}_q(\eta)=(\log(\eta),\eta)'$.
Following Algorithm \ref{algo:b_v2} and using the fact that $E_q[\bm{H}_q(\eta)]=\nabla b(\tau_0,\tau_1)$ as well as approximation (\ref{digammaprox}) for the digamma function $\psi(\cdot)$, the solution of the system $E_q[\bm{H}_q(\eta)]=E_p[\bm{H}_q(\eta)]$ in Step 2.2 results in \(
 \tau_0 =   \tau_1\exp[-(f+Q/2)]\,\,\, \mbox{and} \,\,\,  \tau_1 = (1+\sqrt{1+2Q/3})/(2Q)
\).  The conjugate prior hyperparameters are updated to  
$\tau_1^*=\tau_1+y$ and $\tau_0^*=\tau_0+1$. Step 4.2 is completed by solving the system $E_q[\bm{H}_q(\bm{\lambda})]=E_p[\bm{H}_q(\bm{\lambda})]$, where $p$ is the updated distribution, obtaining:
$f^*=\psi(\tau_1^*)-\log(\tau_0^*)$  and $Q^*=\psi'(\tau_1^*) $. 

The updated moments of the  states $\bm{\theta}$ are trivially obtained by applying normal distribution properties and the predictive distribution for $y_t$ is a Poisson-Gamma($\tau_0, \tau_1,1)$, which can be evaluated once step 2.2 is completed.

\begin{figure}[!htb]
\centering
\includegraphics[width = .975\linewidth]{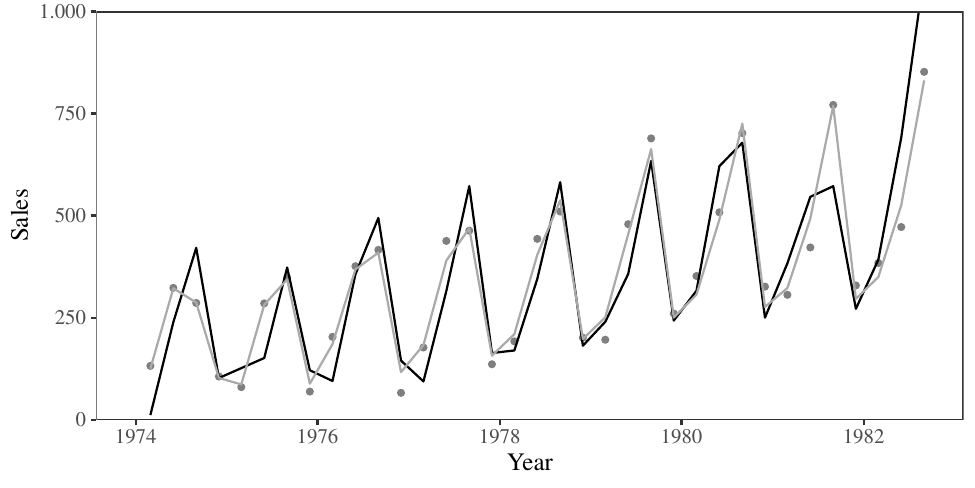}
    \caption{Quarterly sales: observed time  series (solid circles) and the {expected value of the smoothed mean response} function via sequential kDGLM (solid grey line) and local level  method (solid black line).}
    \label{predturkey}
\end{figure}

We apply a Poisson dynamic model to predict a time series of quarterly sales with a growth trend and stochastic seasonal pattern, as detailed in \citet[][Chap. 8, p. 268]{west1997}. True data are shown as solid circles in Figure \ref{predturkey}. We compare this with \citet{west1985dynamic}, using a Poisson log-linear model with linear growth and two pairs of harmonics for a Fourier description of the seasonal pattern. 

Let $    y_t|\eta_t \sim Poisson(\eta_t); \quad     \log(\eta_t)=\bm{F}'_t\bm{\theta}_t ; \quad    \bm{\theta}_t=\bm{G}_t\bm{\theta}_{t-1} +\bm{\omega}_t, \quad \bm{\omega}_t \sim N (\bm{0},\bm{W}_t)
  $,
with $\bm{F}'_t = \PR{1,0,1,0,1,0}; \,\, \bm{G}_t = diag\PR{\bm{G_0},\bm{G_1},\bm{G_2}}
$
and
$$
\bm{G_0} = \left[\begin{array}{cc} 
1 & 1 \\
0 & 1 \\
\end{array}\right],  \bm{G_k} = \left[\begin{array}{cc} 
 \cos(kw) & sin(kw)\\
 -sin(kw) & \cos(kw)\\
 \end{array}\right], w = 2\pi/4 \mbox{ and } k = 1, 2.$$
The prior specification is as follows: $\bm{\theta}_0|D_0 \sim N(m_0=0,C_0=1)$, and we employ a diagonal block discount factor matrix for $\bm{W}_t$ covariance evolution. For the conjugate updating approach of \cite{west1985dynamic}  we specify the same first- and second-order prior moments. Both methods begin with $\bm{\theta}_0|D_0$, updated to the prior $\bm{\theta}_1|D_0$, inducing a predictor's prior structure. 

 Figure \ref{predturkey} exhibits smoothed mean responses, illustrating the proposed method's efficiency in managing predictive structures with multiple dynamic components. Our approach accommodates various stochastic patterns typical in real time series while preserving sequential analysis.
 
We also compare our approach with the local level method proposed by \citet{gamerman2013non}, who introduced Gamma Family Dynamic Models (GFDM) integrating \cite{smith1986non} and \citet{harvey1989time}. GFDM combines a local steady structure (a dynamic level) with  a fixed-coefficient predictor, applicable to models such as Poisson, gamma, Weibull, and normal, as well as others outside the exponential family. Unlike our model, GFDM does not extend beyond the level dynamics. To align with our model's structure, we define regressors $x_1 = t, x_2 = \cos(wt), x_3 = \sin(wt), x_4 = \cos(2wt), x_5 = \sin(2wt)$, where $t = 1, \ldots, 35$ and $w = 2\pi/4$, applying the local level approach of \cite{gamerman2013non} to the following model:
$
y_t|\eta_t \sim Poisson(\eta_t); \; 
\eta_t =\alpha_t\exp(\beta_{1}x_{1t} + \beta_{2}x_{2t} +\beta_{3}x_{3t}+\beta_{4}x_{4t}+\beta_{5}x_{5t}). $
The model is completed with a $Gamma(0.34,0.04)$ prior for $\eta_0$ to standardize prior settings across methods and a $Uniform(0,1)$ prior to manage the increase in the uncertainty associated with the local level approach evolution, impacting dynamic level smoothness \citep{gamerman2013non}. Parameters $\beta_k$, for $k = 1, \ldots, 5$, follow independent $Normal(0, 1)$ priors. The \texttt{NGSSEML} package \citep{santos2021} performed inference with MCMC steps for the static coefficients, and the \texttt{coda} package \citep{plumber} facilitated convergence diagnosis, burn-in, and thinning. Both our method and that of \citet{west1985dynamic} involving conjugate updating provided smoothed estimates and predictions in under one second. Figure \ref{predturkey} illustrates the {expected value of the
smoothed mean response function} obtained from our method and \cite{gamerman2013non}, excluding \cite{west1985dynamic}'s results, which are almost identical to ours. Our method and conjugate updating do not require sampling, resulting in computational efficiency, in contrast to \cite{gamerman2013non}'s local level method, which took 818.803 seconds (effective sample size 1,700), compared to 0.024 second for our approach and \cite{west1985dynamic} (Table \ref{tabcompara1}).

\begin{table}[!htb]
\centering{\small
\label{table_poisson_metric}
\begin{tabular}{@{}lrrr@{}}
\toprule
Metric                    & Local Level & \multicolumn{1}{l}{Conjugate Update} & \multicolumn{1}{l}{kDGLM} \\ \midrule
Mean Absolute Error       & 82.323               & 27.21166     & \textbf{26.86886}                                 \\
Relative Absolute Error   & 0.2661                & 0.1084                & \textbf{0.1077}                              \\
Log Predictive Likelihood & -214.006            & \multicolumn{1}{c}{-}& \textbf{-148.8231}   \\ 
Computational time$^*$ & 818.803s &  \textbf{0.024s} & \textbf{0.024s}\\
\bottomrule
\end{tabular}
\caption{Model comparison based on the smoothed mean response function and several metrics.
\footnotesize{*Time measured using R 4.2.0 running Windows 10 x64 (build 19044) in an Intel(R) Core$^{\mbox{\tiny TM}}$ i7-8700K CPU @ 3.70GHz with 16 GB RAM @ 2400 MHz.}}
\label{tabcompara1}}
\end{table}

\begin{figure}[!htb]
\centering
\includegraphics[width = 1\linewidth]{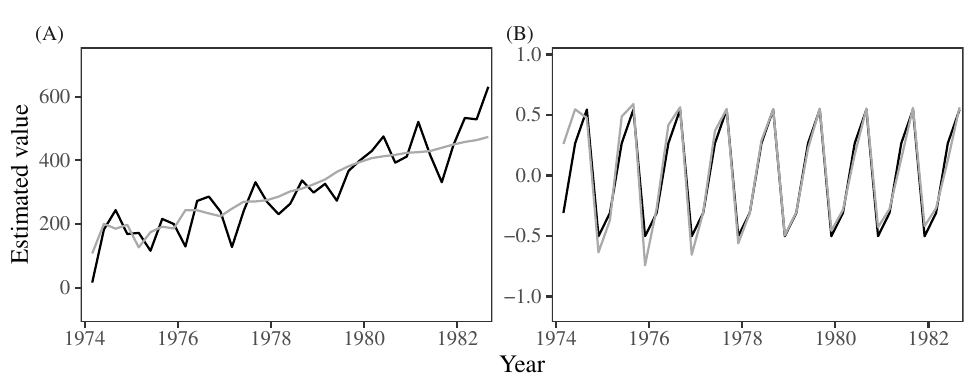}
\caption{Structural decomposition - smoothed mean for seasonal and trend components: kDGLM (gray) and local level (black).}\label{betapois}
\end{figure}

According to comparison of the metrics in Table \ref{tabcompara1}, conjugate updating and our sequential approach are nearly equivalent. The local level model with deterministic seasonality, which is adjustable by the method proposed by \citet{gamerman2013non}, achieved the weakest performance in terms of computational time, as well as regarding fitting and prediction, exhibiting higher uncertainty of smoothed expected values. Unlike \citet{west1985dynamic}, our information geometry approach enables log predictive likelihood evaluation, which is an inherent advantage. The local level approach lacks seasonal component dynamics, as shown in Figure \ref{betapois}, where the trend fit by \citet{gamerman2013non}'s method ``attempts to compensate'' for seasonal inflexibility, showing that our method allows for models that capture structural components with greater precision.

\subsection{Analytical Development of the Proposed Method for Bernoulli Responses}\label{Bernoulli}

{\small
Let  $\eta=\pi$ and $y|\pi \sim Ber(\pi), \, 0<\pi<1$. $\psi(\pi)=\log(\frac{\pi}{1-\pi})$.  The conjugate prior is a beta density  $q(\pi|\bm{\tau}) = exp\{\tau_1 \log(\frac{\pi}{1-\pi})+\tau_0 \log(1-\pi)-[\log(\Gamma(\tau_1+1))+\log(\Gamma(\tau_0-\tau_1+1))-\log(\Gamma(\tau_0+2))]\}$, where $b(\tau_0,\tau_1)=[\log(\Gamma(\tau_1+1))+\log(\Gamma(\tau_0-\tau_1+1))-\log(\Gamma(\tau_0+2))]$. Consider the linear predictor $ \lambda= logit(\pi)=\bm{F}'\bm{\theta}$. 
Following Algorithm \ref{algo:b_v2} :

\begin{itemize}
\item \emph{Step 1}: Given posterior moments for the states, obtain the prior moments of ${\lambda} \sim N({f}, {Q})$.
\item \emph{Step 2.1}:  Obtain the conjugate prior sufficient statistics vector: ${\bm{H}}'_q(\pi)=(\log(\frac{\pi}{1-\pi}),\log(1-\pi))$.
\item \emph{Step 2.2}: Using the fact that $E_q[{\bf H}(\pi)] = \nabla b(\tau_0,\tau_1)$, solve the system $E_q[\bm{H}_q(\pi)]=E_p[\bm{H}_q(\pi)]$, where $p$ is  the prior  density induced by the normal specification for the linear predictors: 
\begin{eqnarray*}
\left\{\begin{array}{ll}
  \psi(\tau_1+1) - \psi(\tau_0-\tau_1+1) &= f\\
  \psi(\tau_0-\tau_1+1)-\psi(\tau_0+2)  &\simeq  \log(\frac{1}{1+e^f}) -\frac{Qe^f}{(1+e^f)^2}.
  \end{array}
  \right.
  \end{eqnarray*}
  A second-order Taylor approximation was applied and the system was solved using the Newton-Raphson method, with negligible computational cost.
 \item \emph{Step 3}: Update the hyperparameters of the conjugate specification:   
$\tau_1^*=\tau_1+y, \tau_0^*=\tau_0+1$. \item \emph{Step 4.1}: Let $q$ denote a normal density for the linear predictor, with sufficient statistics vector: $\bm{H}_q'=({\lambda},\,\,{\lambda^2})$. 
\item \emph{Step 4.2}: Solve the system $E_q[\bm{H}_q(\bm{\lambda})]=E_p[\bm{H}_q(\bm{\lambda})]$, where $p$ is the updated distribution, obtaining:  $f^* \simeq \psi(\tau_1^*+1)-\psi(\tau_0^*-\tau_1^*+1)$\,\, and  \,\, $Q^*=\psi'(\tau_1^*+1)+\psi(\tau_0^*-\tau_1^*+1)$. 
 
\item \emph{Step 5}: The updated moments of the  states $\bm{\theta}$ are trivially obtained, applying normal distribution properties.
\end{itemize}

The predictive distribution for $y_t$ is a binomial-beta($\tau_0, \tau_1,1$), which can be evaluated once step 2.2 is completed. 
}

\subsection{Results for the Simulated Example of the Dynamic Multinomial Model}\label{Mult_simul}

In this section, we evaluate the effect of varying sample sizes $T$ ($T=2,5,  20, 50$), as well as different specifications of the number of allocation categories, $N$, on the approximated inference produced by the sequential kDGLM proposal. We consider synthetic data simulated as described in Section \ref{multinom_simul} and compare the results of our proposed method to those obtained via non-sequential NUTS in Stan.

\begin{figure}[!h]
    \centering
    \includegraphics[width=1\linewidth]{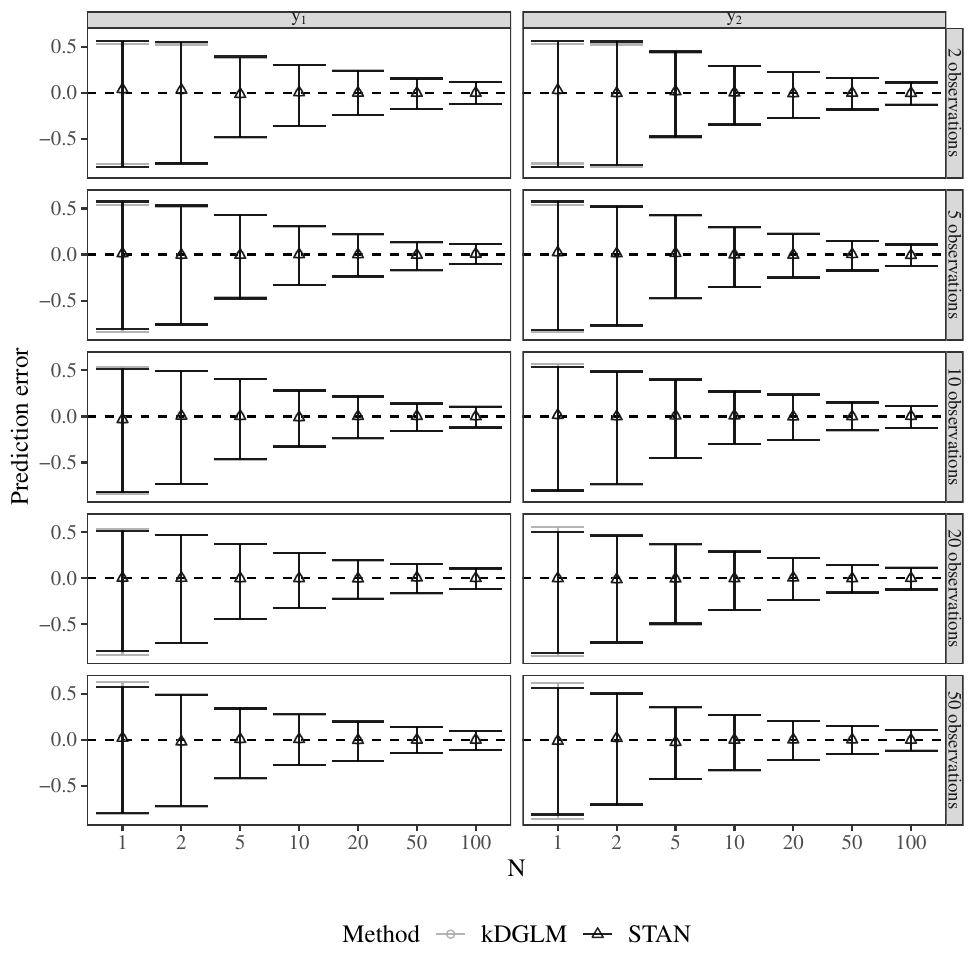}
\caption{The bias for the one-step-ahead prediction for $Y_{1,T+1}$ and $Y_{2,T+1}$. Gray circles and black triangles represent the average bias among all samples, while the error bars represent the $0.975$ and $0.025$ quantiles. To facilitate visualization, the bias is divided by $N$. Gray: Sequential kDGLM. Black: Stan.}
\label{fig:FigSimulMult1}
\end{figure}
\begin{figure}[!h]
    \centering
\includegraphics[width=1\linewidth]{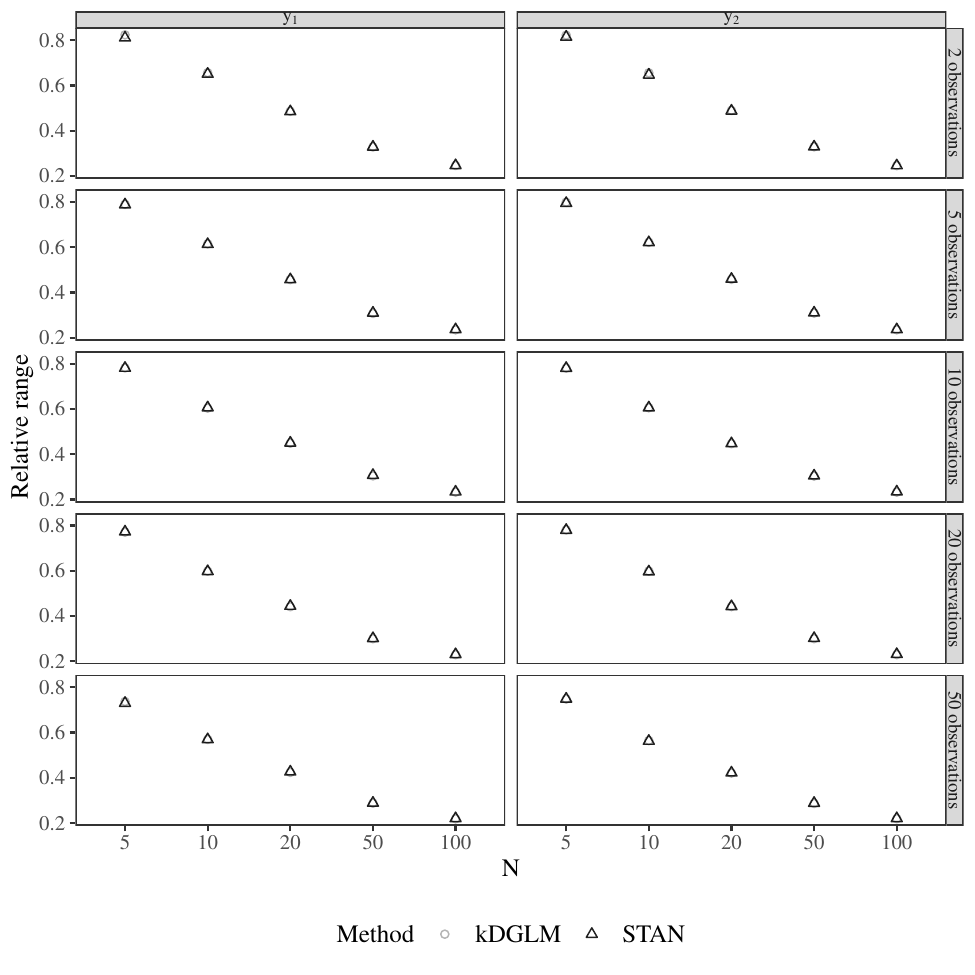}
    \caption{The range of the $95\%$ credibility interval for one-step-ahead prediction for $Y_{1,T+1}$ and $Y_{2,T+1}$. Gray circles: Sequential kDGLM. Black triangles: Stan.}
    \label{fig:FigSimulMult2}
\end{figure}
\begin{figure}[!h]
    \centering
    \includegraphics[width=1\linewidth]{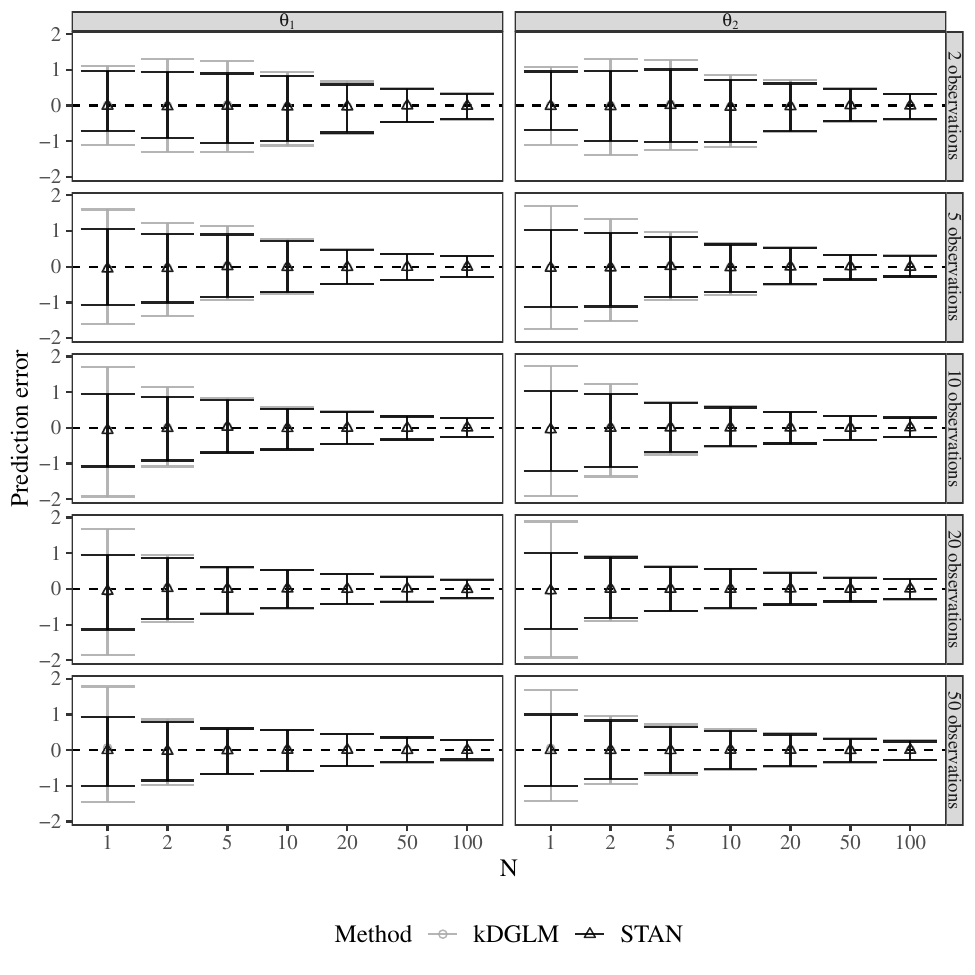}
    \caption{The bias for the latent states at time $T$. Gray circles and black triangles represent the average bias among all samples, while the error bars represent the $0.975$ and $0.025$ quantiles. Gray: Sequential kDGLM. Black: Stan.}
    \label{fig:FigSimulMult3}
\end{figure}
\begin{figure}[!h]
    \centering
    \includegraphics[width=1\linewidth]{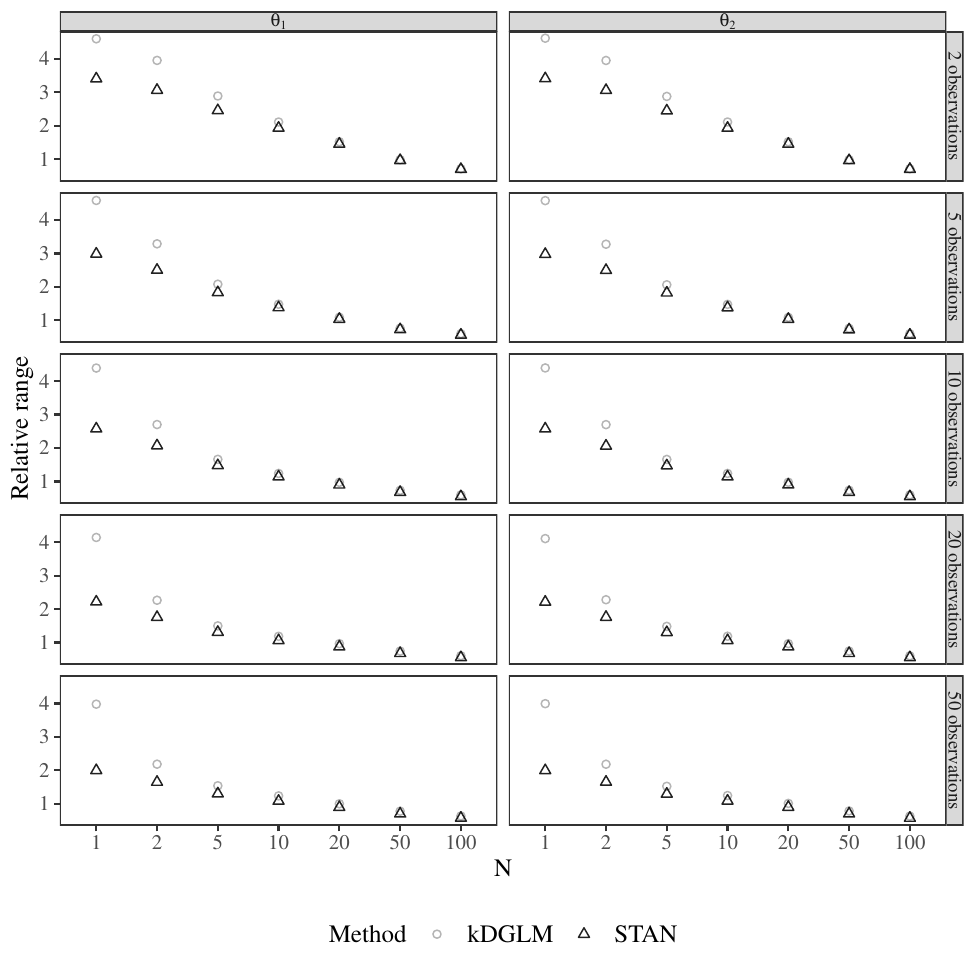}
    \caption{The range of the $95\%$ credibility interval for the latent states at at time $T$. Gray: Sequential kDGLM. Black: Stan.}
    \label{fig:FigSimulMult4}
\end{figure}

\clearpage

\subsection{Multinomial Model: Obtaining the Probabilities of Hospital Admission for Each Age Group\label{Offset}}

{\small As seen in Subsection \ref{case}, for a multinomial response on $k$ categories where the $k-$th category is used as a reference group, the following predictive structure is specified:

$$
\lambda_{jt}=  \log\PC{\frac{\eta_{jt}}{\eta_{k,,t}}}= \bm{F}_{jt}'\bm{\theta_t}, \quad j=1,\ldots, k-1.
$$

Let $E_{jt}$ denote the exposure of group $j$; $E_{1:k,t}=\sum_{j=1}^ kE_{jt}$ denote the total exposure of all age groups, and consider the following events:

$G_{jt}:$ allocation of an individual to age group $j$, $j=1,\ldots, k-1$; at time $t$, t=1, 2, \ldots;

$H_t$:  hospital admission of an individual at time $t$ , t=1, 2, \ldots.

Note that the probability of allocation to age group $j$ and time $t$, conditionally on having been hospitalized, is:
\begin{equation*}
\eta_{jt}=\mathbb{P}(G_{jt}|H_t)= \frac{ \mathbb{P}(H_t|G_{jt})\mathbb{P}(G_{jt}) }{\mathbb{P}(H_t)} \Rightarrow \frac{\mathbb{P}(H_t|G_{jt})}{\mathbb{P}(H_t|G_{kt})}=\frac{\mathbb{P}(G_{jt}|H_t)}{\mathbb{P}(G_{kt}|H_t)}\frac{E_{kt}}{E_{jt}}.
\end{equation*}
Then it follows that:

$$
\begin{aligned}
\log\left\{\frac{\mathbb{P}(H_t|G_{jt})}{\mathbb{P}(H_t|G_{kt})
}\right\}=&\log\left\{\frac{\mathbb{P}(G_{jt}|H_t)}{\mathbb{P}(G_{kt}|H_t)}\right\}-\log\left\{\frac{E_{jt}}{E_{kt}}\right\}\\
=& \bm{F}_{jt}'\bm{\theta}_t-\log\left\{\frac{E_{jt}}{E_{kt}}\right\}, \quad j=1,...,k-1.
\end{aligned}
$$

Thus, estimates of $\mathbb{P}(H_t|G_{jt})$ are naturally obtained if  $\log\left\{\frac{E_{jt}}{E_{kt}}\right\}$ is introduced as an offset term in  the dynamic predictor of category $j$, $j=1,...,k-1$.}

\subsection{Analytical Development of the Proposed Method for the Normal Family with unknown mean and precision \label{Proof}}

This section provides a detailed description of the analytical developments involved in updating information for the normal distribution within the kDGLM sequential method. This is an instance presented in this work in which the application of the suggested methodology is not straightforward. The outlined procedure is executed at each time point $t$ during the information updating process.

    Consider the general formulation and parametrization defined in Section \ref{volat_normal} for the normal model with dynamic predictive structure for the mean and precision. Since $\bm{\lambda} \sim N_2(\bm{f}, \bm{Q})$,  a system of equations in $\tau$ remains to be solved, that is, $E_q[\bm{H}_q(\mu, \phi)]=E_p[\bm{H}_q(\mu,\phi)]$, with $q$ denoting a normal-gamma prior distribution and $p$, the density implied by the joint normality of the linear predictors for the mean ad log-precision. Using the facts that: i) $E_q[\bm{H}_q(\mu, \phi)]=\nabla b(\tau_1,\tau_2,\tau_3,\tau_0)$; ii) $E_p[\bm{H}_q(\mu,\phi)]=E_{p(\phi)}[E_{p(\mu|\phi)}\bm{H}_q(\mu,\phi)]$ which can be obtained by normal conditioning properties; iii) under $p$, $\phi \sim LN(f_2,q_2)$ and $\ln \phi \sim N(f_2,q_2) $, results in the system:
\begin{eqnarray}
    \left\{ \begin{array}{lll}
    \frac{(2\tau_0+1)\tau_2^2}{2\tau_1\tau_2^2-8\tau_1^2\tau_3}-\frac{1}{2\tau_1}&=& \exp(f_2+q_2/2)[(f_1+q_{12})^2+q_1]\\
     \frac{-(2\tau_0+1)\tau_2}{\tau_2^2-4\tau_1\tau_3}&=&\exp(f_2+q_2/2)(f_1+q_{12})\\
     \frac{4\tau_1(\tau_0+1/2)}{\tau_2^2-4\tau_1\tau_3}&=&\exp(f_2+q_2/2)\\
     \psi(\tau_0+1/2)-\ln\PC{\frac{\tau_2^2}{4\tau_1}-\tau_3}&=&f_2
      \end{array}\right.,
      \label{eqs_ida_normal}
    \end{eqnarray}
with $\psi(\cdot)$ denoting the digamma function.  This system can be analytically solved, as follows: applying  the parametrization:

\begin{equation}
   c_0=-2\tau_1, \quad 
    \mu_0=-\frac{\tau_2}{2\tau_1}, \quad
    \frac{d_0}{2}= \frac{\tau_2^2}{4\tau_1}-\tau_3 , \quad
    \frac{n_0}{2}= \tau_0+1/2,
    \label{reparam}
\end{equation}
to  the third equation in  (\ref{eqs_ida_normal}), it follows that $n_0/d_0=\exp(f_2+q_2/2)$. Multiplying the third equation by $-\frac{2\tau_2}{4\tau_1}=\mu_0$ we obtain the second equation in  (\ref{eqs_ida_normal}) resulting in $\mu_0=f_1+q_{12}$. Doing some algebra in the first equation of (\ref{eqs_ida_normal}), it follows  that $c_0=[q_1\exp(f_2+q_2/2)]^{-1}$. Finally, applying the approximation $\psi(u)\approx \ln(u)-1/(2u)$ to the fourth equation of the system, results in $n_0=2/q_2$. Applying the reparametrization in (\ref{reparam}) we obtain the following closed analytical expressions:
\begin{eqnarray*}
    \left\{ \begin{array}{lll}
    \tau_0&=&1/q_2-1/2\\
     \tau_1&=&-[2q_1\exp(f_2+q_2/2)]^{-1}\\
     \tau_2&=&(f_1+q_{12})[q_1\exp(f_2+q_2/2)]^{-1}\\
      \tau_3&=&-(f_1+q_{12})^2[2q_1\exp(f_2+q_2/2)]^{-1}-[q_2\exp(f_2+q_2/2)]^{-1}. 
      \end{array}\right..
      \label{eqs_ida_normal_final_appendix}
    \end{eqnarray*}

After observing $\bm{y}$, we obtain  $\bm{\tau}^*=(\tau_1^*,\tau_2^*,\tau_3^*, \tau_0^*)=(\tau_1-1/2,\tau_2+y,\tau_3-y^2/2,\tau_0+1/2)$, the updated canonical parameters of a normal-gamma posterior density for $(\mu, \phi)$, and need to evaluate the parameters $(\bm{f}^*, \bm{Q}^*)$ of the posterior distribution of the linear predictors that are compatible with them, solving: $E_q[\bm{H}_q]=E_p[\bm{H}_q]$, now considering that $p$ is a normal-gamma density for $(\mu, \phi)$ and $q$ is a bivariate normal for $(\mu, \ln \phi)$. $\bm{H}_q'=(\bm{\lambda},\bm{\lambda}\bm{\lambda}')$,   so the following system is trivially solved:
\begin{equation*}
    \left\{ \begin{array}{lll}
    f_1^*&=& -\frac{\tau_2^*}{2\tau_1^*}\nonumber\\
    f_2^*&=& \psi(\tau_0^*+1/2)-\ln\PC{\frac{(\tau_2^*)^2}{4\tau_1^*}-\tau_3^*}\nonumber\\
     Q_{11}^*&=&\frac{\tau^{*2}_2-4\tau_1\tau_3}{4\tau_0\tau_1-2\tau_1}\nonumber\\
     Q_{12}^*&=&0\nonumber\\
     Q_{22}^*&=&\psi'(\tau_0^*+1/2).
 \end{array}\right.
    \end{equation*}

Notice that, no matter the values of $\tau_0^*,\tau_1^*,\tau_2^*$ and $\tau_3^*$, $Q^*_{12}=0$, since the parameters that minimize the KL divergence from $p$ to $q$ must satisfy $Cov_p[\mu,\ln(\phi)]=Cov_q[\mu,\ln(\phi)]$ and, as a property of the normal-gamma distribution, we have that $(\mu,\phi)$ are uncorrelated (although not independent), consequently, it can be shown that $(\mu,\ln \phi)$ are also uncorrelated.

Using the reparametrization 
\(    c^*_0=-2\tau^*_1, \quad 
    \mu^*_0=-\frac{\tau^*_2}{2\tau^*_1}, \quad
    \frac{d^*_0}{2}= \frac{\tau^{*2}_2}{4\tau^*_1}-\tau^*_3 \quad \mbox{and} \quad
    \frac{n^*_0}{2}= \tau^*_0+1/2
  \), we obtain:

\begin{equation*}
    \left\{ \begin{array}{lll}
    f_1^*&=& \mu^*_0\nonumber\\
    f_2^*&=& \psi(n^*_0/2)-\ln\PC{d^*_0/2}\nonumber\\
     Q_{11}^*&=&\frac{d_0^*/2}{c^*_0(n^*_0/2-1)}\nonumber\\
     Q_{12}^*&=&0\nonumber\\
     Q_{22}^*&=&\psi'(n_0^*/2).
 \end{array}\right.
    \end{equation*}

An important point to discuss is the equation that defines $Q_{11}^*$. Specifically, notice that $Q_{11}^*$ is not defined for $n_0^*\leq 2$. Indeed, by the Projection Theorem, the parameters that minimize the KL divergence from $p$ to $q$ must satisfy $Var_p[\mu]=Var_q[\mu]$, but in the conjugated posterior distribution $\mu| \phi \sim N(\mu_0^*,(c_0^*\phi)^{-1})$ and $\phi \sim \mathcal{G}(n_0^*/2,d_0^*/2)$, which implies that $\mu \sim t\left(n_0^*,\mu_0^*,\frac{d_0^*/2}{c^*_0n^*_0/2}\right)$, so that when $n_0^* \leq 2$, $Var_p[\mu]=+\infty$, and the system $E_q[\bm{H}_q]=E_p[\bm{H}_q]$ has no valid solutions. Indeed, one can always reduce the KL divergence from $p$ to $q$ by increasing $Q_{11}^*$.
This situation is quite undesirable since it compromises the use of our proposed methodology. Yet, we can restrict the parameter space for the normal posterior distribution to guarantee that, inside that restricted space, we always have a minimum for the divergence from $p$ to $q$. A natural restriction for the parameter space is to set $Q_{11}^*=\frac{d_0^*/2}{c_0^*n^*_0/2}$, since:

\begin{itemize}
    \item This restriction guarantees that the scale parameter for the marginal distribution of $\mu$ is identical in both the normal and the conjugated distribution.
    \item for a large value of $n^*_0$ (which we expect to have for a reasonable sample size), $\frac{d_0^*/2}{c_0^*n^*_0/2} \approx \frac{d_0^*/2}{c_0^*(n^*_0/2-1)}$, i.e., for large values of $n^*_0$ this restriction has no significant effect, in the sense that the optimum in the restricted space will be very close to the global optimum.
    \item $Q_{11}^*=\frac{d_0^*/2}{c_0^*n^*_0/2}$ becomes numerically well behaved for any possible value of $n^*_0$, since, after updating our knowledge of $\mu$, it is guaranteed that  $n^*_0 \geq 1$, avoiding a division by values close to zero.
\end{itemize}

\subsection{Comparison Between kDGLM and NUTS via Stan -- Stochastic Volatility Example}\label{S_Stan}

Section \ref{artificial} presents a comparison between the results obtained using the sequential kDGLM method proposed in this work and those from a NUTS scheme implemented via Stan. The comparison relies on synthetic, artificially generated data. As discussed, the differences observed in the inference produced by the two methods are largely attributable to the linearization employed for sequential inference for the autoregressive parameter 
$\gamma$
 in the kDGLM method.

\begin{figure}[!h]
\centering
\includegraphics[width = 1.065\linewidth]{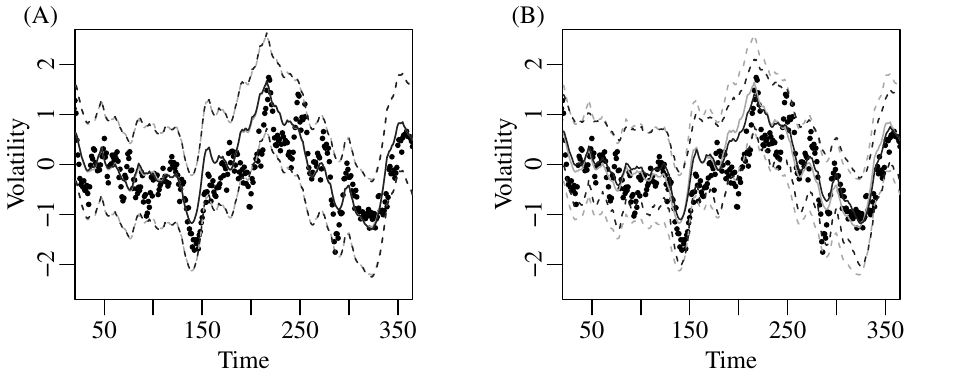}
    \caption{
Comparison of smoothed means (solid lines) and credible intervals (dashed lines) for volatility estimation from synthetic data, with fixed autoregressive parameter $\gamma$. The solid circles indicate theoretical volatilities. (A): results of the proposed sequential approach for normal (gray) and gamma (black) formulations. (B): comparison of normal formulation using the sequential approach (gray) and NUTS via Stan (black).}
    \label{fig:compara_simul_supl}
\end{figure}

In Figure \ref{fig:compara_simul_supl}, we display the inference results under both methods, kDGLM and NUTS, with the parameter 
$\gamma$
 fixed at its true value. As shown, the two methods yield nearly identical results. Therefore, for models where the evolution matrix 
$\mathbf{G}$
 is known, the kDGLM method is expected to produce inferences closely aligned with those generated by Markov Chain Monte Carlo schemes.

\clearpage

\bibliographystyle{apalike}
\bibliography{references}

\end{document}